\documentclass[11pt]{article}

\usepackage{geometry}
\usepackage[utf8]{inputenc}
\usepackage{enumitem,amssymb}
\usepackage{graphicx,amsmath}
\usepackage{authblk}
\usepackage{url}
\usepackage{multirow}
\usepackage{textpos}

\usepackage[
 colorlinks=true,    % false: boxed links; true: colored links 
 linkcolor=blue,     % color of internal links            
 citecolor=blue,     % color of links to bibliography
 filecolor=blue,  % color of file links 
 urlcolor=blue,      % color of external link
 final=true
]{hyperref}
\usepackage[numbers,sort&compress]{natbib}
\usepackage{enumitem}
\usepackage[symbol]{footmisc}
\usepackage[dvipsnames]{xcolor}
\usepackage{tabularx,booktabs,caption}

\newcommand{\eehz}{$e^+ e^- \rightarrow h Z$}

\setenumerate{itemsep=0mm}

\begin{document}

\thispagestyle{empty}

\title{\large ~~~~~~~~~~~~~~~~~~~~~~~~~~~~~~~~~~~~~~~~~~~~~~~~~~~~~~~~~~~~~~~~~~ \textsc{Snowmass 2021 } KIAS--P22014\\
\LARGE \textbf{Global fit of 2HDM with future collider results}}

\author[a]{Ankit Beniwal}
\author[b]{Filip Rajec}
\author[c]{Markus~Tobias~Prim}
\author[d,e]{Pat Scott}
\author[f]{Wei Su*}
\author[b]{Martin White}
\author[b]{Anthony G. Williams}
\author[b]{Alex Woodcock}

\affil[a]{\small Theoretical Particle Physics and Cosmology Group, Department of Physics, King’s College London, Strand, London, WC2R 2LS, UK}
\affil[b]{\small ARC Centre of Excellence for Dark Matter Particle Physics, Department of Physics, University of Adelaide, South Australia 5005, Australia}
\affil[c]{Physikalisches Institut der Rheinischen Friedrich-Wilhelms-Universit\"at Bonn, 53115 Bonn, Germany}
\affil[d]{School of Mathematics and Physics, The University of Queensland, St. Lucia, Brisbane, QLD
4072, Australia}
\affil[e]{Department of Physics, Imperial College London, Blackett Laboratory, Prince Consort Road, London SW7 2AZ, UK}
\affil[f]{Korea Institute for Advanced Study, Seoul 02455, Korea}

\maketitle

% \normalsize
\footnotetext[1]{Email\,address:\,\url{weisu@kias.re.kr}}
\noindent {\large \bf Thematic Areas:}  %(check all that apply $\square$/$\blacksquare$)

\noindent $\blacksquare$ (EF01) EW Physics: Higgs Boson properties and couplings \\
\noindent $\blacksquare$ (EF02) EW Physics: Higgs Boson as a portal to new physics \\
\noindent $\blacksquare$ (EF04) EW Precision Physics and constraining new physics \\
\noindent $\blacksquare$ (EF08) BSM: Model specific explorations \\
\noindent $\blacksquare$ (TF07) Collider phenomenology \\

% \noindent {\large \bf Abstract:} (maximum 200 words)
\begin{abstract}
\large 
In this work, we summarize a global fit study of Type-II two Higgs doublet models (2HDM), and explore the impact of future SM-like Higgs and $Z$-pole precision measurements on the allowed parameter space. % 
The work is based on the study results of a global fit of 2HDMs with the tool \textsf{GAMBIT}, utilising various current constraints including theoretical constraints (unitarity, perturbativity and vacuum stability), Higgs searches at colliders, electroweak physics and flavour constraints.  We further investigate the ability of future facilities, such as the HL-LHC,  CEPC, ILC and FCC-ee to explore the 2HDM parameter space.
\end{abstract}

% \clearpage

\large 

\section{Introduction}
The discovery of a Standard Model (SM)-like Higgs boson at the Large Hadron Collider (LHC) set a milestone for high energy physics, confirming the self-consistency of the SM. At the same time there are also various unsolved mysteries, such as the source of dark matter, the origin of the baryon asymmetry of the universe, and the muon $g$-2 anomaly. Models with extended Higgs sectors provide promising solutions to these problems.

As one of the simplest such frameworks, the Two Higgs Doublet Model (2HDM) is embedded in various models with extended Higgs sectors, such as the Minimal Supersymmetric Standard Model, and gauge extensions (such as the Left-Right symmetric model). After electroweak symmetry breaking (EWSB), the general CP-conserving 2HDM can generate five physical eigenstates: the observed 125 GeV CP-even neutral scalar $h$, an additional CP-even neutral scalar $H$, one CP-odd Higgs boson $A$, and a pair of charged Higgs bosons $H^\pm$~\cite{Branco:2011iw}.  Exploring the properties of 2HDMs with various experimental constraints can help us understand the new physics potential of a broad class of BSM scenarios.

In this paper, we present preliminary results of a forthcoming global fit of the $Z_2$-Yukawa symmetric, Type-II 2HDM \cite{gambit_Type-II_2HDM}. This analysis is carried out using the open-source tool \textsf{GAMBIT} \cite{GAMBIT:2017yxo} (Global and Modular beyond-Standard Model Inference Tool) %\Ankit{Previously Ref.~[2] pointed to \textsf{GAMBIT} Higgs portal study. I've corrected this to the \textsf{GAMBIT} v1 paper now}. 
\textsf{GAMBIT} is compatible with both the Bayesian and frequentist statistical frameworks, and we here focus on frequentist results obtained with the \texttt{Diver}~\cite{Workgroup:2017htr} implementation of the differential evolution algorithm. We investigate the effect of theoretical constraints (unitarity, perturbativity and vacuum stability), Higgs searches at colliders, electroweak physics and flavour constraints individually, as well as displaying the final results with all constraints. We also investigate the impact on the allowed parameter space of a series of future collider measurements, by reweighting the likelihoods of the \textsf{GAMBIT} samples outside of the \textsf{GAMBIT} framework 
%{\bf MJW: Wei - is that true? I made it up assuming it is what you did!} Yes

Our paper is organised as follows. Section~\ref{sec:input} provides the details of our assumed future collider measurements. Section~\ref{sec:2hdm} summarises the general 2HDM and our results are presented in Section~\ref{sec:results}.

\section{Higgs precision measurements at future lepton colliders}\label{sec:input}
At future lepton colliders, the dominant channel to measure the properties of the Higgs boson is the Higgsstrahlung process, \eehz, at center of mass energies $(\sqrt{s})$ of around 240$-$250\,GeV.  Due to the nature of lepton colliders, both the inclusive cross section, $\sigma(hZ)$, and the exclusive $\sigma(hZ)\times {\rm BR}$ values for different Higgs decay modes, can be measured with remarkable precision.  The invisible decay width of the Higgs can also be very well constrained. In addition, the cross section for a Higgs production via the $WW$ fusion process grows with energy.~While it cannot be measured very well at 240$-$250\,GeV, at higher center of mass energies (in particular, at linear colliders), such a fusion process becomes significantly more important and can provide crucial complementary information. For $\sqrt{s}>500$ GeV, $tth$ production can also be investigated.

%%%%%%%%%%%%%%%%%%%%%%%%%%%

\begin{table}[tb]
 \begin{center}
  \begin{tabular}{|l|r|r|r|r|r|r|r|r|r|r|}
   \hline
   collider& \multicolumn{1}{c|}{CEPC}& \multicolumn{3}{c|}{FCC-ee}&\multicolumn{5}{c|}{ILC} \\
   \hline
   $\sqrt{s}$     &  $\text{240\,GeV} $ &  $\text{240\,GeV}$	&\multicolumn{2}{c|}{$\text{365\,GeV}$}  &  \text{250\,GeV}  &
   \multicolumn{2}{c|}{\text{350\,GeV}}  & \multicolumn{2}{c|}{\text{500\,GeV}} \\
   %\hline
   $\int{\mathcal{L}}dt $     &  $\text{5.6 ab}^{-1} $ &  $\text{5 ab}^{-1}$	&
   \multicolumn{2}{c|}{$\text{1.5 ab}^{-1}$}      &  $\text{2 ab}^{-1} $  &
   \multicolumn{2}{c|}{$\text{200 fb}^{-1}$}  & \multicolumn{2}{c|}{$\text{4 ab}^{-1}$} \\
   \hline
    \hline
production& $Zh$  & $Zh$   & $Zh$   &$\nu\bar{\nu}h$     & $Zh$      & $Zh$     & $\nu\bar{\nu}h$     & $Zh$     & $\nu\bar{\nu}h$  \\
   \hline
  $\Delta \sigma / \sigma$ & 0.5\%  & 0.5\%& 0.9\% &$-$ & 0.71\% & 2.0\% & $-$ & 1.05 & $-$ \\ \hline \hline
   decay & \multicolumn{9}{c|}{$\Delta (\sigma \cdot BR) / (\sigma \cdot BR)$}  \\
  \hline
   $h \to b\bar{b}$              &  0.27\%               & 0.3\%       			  &  0.5\%			& 0.9\%		           &  0.46\%    & 1.7\%         & 2.0\%                  & 0.63\%    & 0.23\%         \\

   $h \to c\bar{c}$              & 3.3\%                   & 2.2\%                   &6.5\%			&10\%					& 2.9\%          & 12.3\%    & 21.2\%                   & 4.5\%     & 2.2\%                    \\

   $h \to gg$                    & 1.3\%                   & 1.9\%                  & 3.5\%           & 4.5\%    & 2.5\%           & 9.4\%    & 8.6\%                  & 3.8\%     & 1.5\%                  \\

   $h \to WW^*$                  & 1.0\%                   & 1.2\%                   & 2.6\%          & 3.0\%    & 1.6\%          & 6.3\%    & 6.4\%                  & 1.9\%    & 0.85\%               \\

   $h \to \tau^+\tau^-$         & 0.8\%                   & 0.9\%                   &1.8\%           &8.0\%          &1.1\%           &4.5\%          & 17.9\%                  & 1.5\%    & 2.5\%               \\

   $h \to ZZ^*$                  & 5.1\%                   & 4.4\%                   & 12\%        & 10\%     & 6.4\%        & 28.0\%     & 22.4\%                   & 8.8\%     & 3.0\%                \\

   $h \to \gamma\gamma$          & 6.8\%                   & 9.0\%                   & 18\%     & 22\%     & 12.0\%     & 43.6\%     &50.3\%                   & 12.0\%   &6.8\% \\

   $h \to \mu^+\mu^-$           & 17\%                   & 19\%                  & 40\%        & $-$     & 25.5\%        & 97.3\%     & 178.9\%                  & 30.0\%     & 25.0\%                    \\
   \hline
    $(\nu\bar\nu)h \to b\bar{b}$  & 2.8\%       &   3.1\%      & $-$  & $-$ &       3.7\% & $-$  & $-$  & $-$  & $-$   \\
    \hline
  \end{tabular}
  \caption{Estimated statistical precisions for Higgs measurements obtained at  the proposed CEPC program with 5.6 ab$^{-1}$ integrated luminosity~\cite{CEPCStudyGroup:2018ghi,CEPCPhysics-DetectorStudyGroup:2019wir}, FCC-ee program with 5 ab$^{-1}$ integrated luminosity~\cite{Abada:2019lih,Abada:2019zxq},  and  ILC with various center-of-mass energies~\cite{Bambade:2019fyw}. 
   }
\label{tab:mu_precision}
  \end{center}
\end{table}
When investigating the impact of future facilities, our study makes use of the following scenarios of various machines (in terms of the center of mass energy and the corresponding integrated luminosity), as well as the estimated precision of relevant Higgs measurements:
\begin{itemize}

    \item  {\bf CEPC~~}  According to the preCDR \cite{CEPC-SPPCStudyGroup:2015csa}, CEPC plans to collect 5 ab$^{-1}$ data points at 240\,GeV.  The estimated precision of measurements for the Higgsstrahlung process \eehz~with various final states, as well as the $WW$ fusion process with Higgs decaying to bottom pairs ($e^+ e^- \rightarrow \nu \bar{\nu} h$, $h\to b\bar{b}$) are summarized in \autoref{tab:mu_precision}.  As systematic uncertainties of the Higgs measurements can be reduced using $Z$-pole calibration, they are assumed to be much smaller than the statistical uncertainties and are therefore neglected.

     \item  {\bf FCC-ee~~}  The FCC-ee CDR is finished in 2018 \cite{FCC:2018evy,Abada:2019lih,Abada:2019zxq}.At the current moment, the white paper proposes total luminosities of $\text{5 ab}^{-1}$ at 240\,GeV and $1.5 \text{ab}^{-1}$ at 350\,GeV. The estimated precision of \eehz~measurements at 240\,GeV, as well as $h\rightarrow b\bar{b}$ channel in $WW$ fusion are listed in \autoref{tab:mu_precision}. In addition, the  cross sections of vector boson fusion processes for the Higgs production ($WW,ZZ\to h$) grow with the center of mass energy logarithmically.  While their rates are still rather small at 240-250 GeV, at higher energies such as 350 GeV,  such fusion processes become significantly more important and can provide crucial complementary information. 

    \item  {\bf ILC~~}  The proposed run scenarios in the ILC TDR  \cite{Baer:2013cma} have been updated in recent documents \cite{Fujii:2015jha, Barklow:2015tja}, which suggested that the ILC could collect $2 \text{ab}^{-1}$ data points at 250\,GeV, $200 \text{fb}^{-1}$ at 350\,GeV, and $4 \text{ab}^{-1}$ at 500\,GeV.  However, the estimation of signal strengths, as summarized in Ref.~\cite{Barklow:2015tja}, are only available for smaller benchmark luminosities for which the full detector studies are performed. We take these estimations and scale them up to the current run scenarios, assuming statistical uncertainties dominate~\cite{Asner:2013psa}; these are summarized in \autoref{tab:mu_precision}.~Such scaling provides a reasonable approximation as long as the luminosities are not excessively large and the systematic uncertainties are under control.

\end{itemize}
With large center of mass energies up to 3\,TeV, CLIC is also able to measure the Higgs properties very well through the $WW$ fusion process \cite{CLIC:2016zwp, Abramowicz:2016zbo}. On the other hand, with its extensive coverage of energy scales, the primary goal of CLIC is to directly search for new particles, in particular, the ones coupled to SM particles only through electroweak interactions.
% CLIC is advantageous in the possibility of directly producing   new particles that modifies Higgs properties at low energy, and a large span of observables at different scale. Consequently, a study on the CLIC physics potential for various models would require additional considerations beyond Higgs precision physics.
 A comprehensive study of the CLIC physics potential including both the direct and indirect searches of new physics is beyond the scope of this paper.

In our global fit to the Higgs measurements, we only include the rate information for the Higgsstrahlung as well as the $WW$ fusion process.  
Electroweak (EW) precision measurements at the $Z$-pole also impose strong constraints on new physics~\cite{Gori:2015nqa,Su:2016ghg}.  The current constraints from the Large Electron Positron (LEP) collider can be significantly improved by a $Z$-pole run at any of the future lepton colliders. While these constraints are not explicitly considered in our study, we do restrict ourselves to models with suppressed EW precision corrections ({\it e.g.,} by imposing custodial symmetries) such that these constraints are automatically satisfied.

It is also important to study the reach of the future High Luminosity LHC (HL-LHC)~\cite{Cepeda:2019klc,ATLAS:2019mfr,ATL-PHYS-PUB-2014-016}. Current LHC Higgs measurements are included via the \textsf{GAMBIT} interfaces to \textsf{HiggsBounds-5.3.2} \cite{Bechtle:2008jh,Bechtle:2015pma} and \textsf{HiggsSignals-2.2.3} \cite{Bechtle:2013xfa}. For the future HL-LHC, we take their designed precision measurements to construct likelihood, which will be introduced in detail later.
% {\bf MJW: Need to give versions of these codes, and list which data are included. Also, this paragraph does not actually say what we did for the HL-LHC}. 
%
% The detailed inputs are listed in \autoref{app:lhcinput}, with the LHC Run-I results in \autoref{tab:HIGGS_datarun12} and the ATLAS projections for LHC $300\infb$ and $3000\infb$ summarized in \autoref{tab:HIGGS_datarun3000}.

% Nowadays there have been many Higgs factory proposals, including the CEPC in China~\cite{CEPC-SPPCStudyGroup:2015csa,CEPCStudyGroup:2018ghi}, the FCC-ee at CERN~\cite{Abada:2019lih,Abada:2019zxq,Gomez-Ceballos:2013zzn,fccpara,fccplan,Blondel:2019yqr} and the ILC in Japan~\cite{Baer:2013cma,Bambade:2019fyw,Fujii:2019zll,Fujii:2020pxe}, which can reach sub-percentage precision on the Higgs properties. In addition, it is foreseen to run these colliders at the Z-pole. These machines will provide improved measurements of SM parameters, demonstrating an impressive potential for future precision measurements of SM Higgs observables~\cite{Chen:2018shg,Chen:2019pkq}.

Based on these analyses, we propose a global fit study of the 2HDM with hypothetical data from Higgs and Z pole precision measurements at future HL-LHC and Higgs factories.

\section{2HDM and study strategy }
\label{sec:2hdm}
The general 2HDM has two ${\rm SU}(2)_L$ scalar doublets $\Phi_i\ (i=1,2)$ with hyper-charge $Y=+1/2$,
\begin{equation}
\Phi_{i}=\begin{pmatrix}
  \phi_i^{+}    \\
  (v_i+\phi^{0}_i+iG_i)/\sqrt{2}
\end{pmatrix}\,,
\end{equation}
where $v_i\ (i=1,\,2)$ are the vacuum expectation values (VEVs) of the two doublets after EWSB with $v_1^2+v_2^2 = v^2 = (246\ {\rm GeV})^2$ and $\tan \beta = v_2/v_1$. 

The 2HDM Lagrangian for the Higgs sector can be written as
\begin{equation}\label{equ:Lall}
\mathcal{L}=\sum_i |D_{\mu} \Phi_i|^2 - V(\Phi_1, \Phi_2) + \mathcal{L}_{\rm Yuk}\,,
\end{equation}
with a Higgs potential of
\begin{align}\label{eq:L_2HDM}
 V(\Phi_1, \Phi_2) &= m_{11}^2\Phi_1^\dag \Phi_1 + m_{22}^2\Phi_2^\dag \Phi_2 -m_{12}^2(\Phi_1^\dag \Phi_2+ h.c.) 
 + \frac{\lambda_1}{2}(\Phi_1^\dag \Phi_1)^2 + \frac{\lambda_2}{2}(\Phi_2^\dag \Phi_2)^2  \nonumber \\
 &\hspace{5mm} + \lambda_3(\Phi_1^\dag \Phi_1)(\Phi_2^\dag \Phi_2)+\lambda_4(\Phi_1^\dag \Phi_2)(\Phi_2^\dag \Phi_1) 
 + \frac{\lambda_5}{2}   \Big[ (\Phi_1^\dag \Phi_2)^2 + h.c.\Big]\,,
\end{align}
where we have assumed $CP$ conservation and a soft $\mathbb{Z}_2$ symmetry breaking term $m_{12}^2$. The physical degrees of freedom after EWSB are a pair of singly-charged Higgs bosons $H^\pm$, a CP-odd Higgs boson $A$ and two CP-even Higgs bosons $h$ and $H$. Here we take $h$ as the observed Higgs boson with a mass of 125 GeV.  Our current study foucs on the Type-II 2HDM, where one Higgs doublet couples to up-type quarks, and the other Higgs doublet couples to down-type quarks and leptons.

For the study of current constraints, we refer to the forthcoming work of the \textsf{GAMBIT} collaboration~\cite{gambit_Type-II_2HDM}, which includes the latest theoretical constraints (unitarity, perturbativity and vacuum stability), Higgs searches at colliders, electroweak physics and flavour constraints.  To further investigate the precise measurement constrains of future colliders, such as HL-LHC,  CEPC, ILC and FCC-ee, we define new likelihoods for the proposed Higgs factories as 
\begin{equation}
-2\ln\mathcal{L}_{\rm{Future}} = \frac{(m_h-m_h^{\rm obs})^2}{\delta_{m_h}^2} 
+ \sum_i \frac{(\mu_i-\mu_i^{\rm obs})^2}{\sigma_{\mu_i}^2}+\sum_{ij} ( X_i - \hat X_i) (\sigma^2)_{ij}^{-1} ( X_j - \hat{X}_j )\,,
\end{equation}
Here $\mu_i^{\rm BSM} \equiv (\sigma \cdot \rm{BR})_{\rm BSM}/(\sigma \cdot \rm{BR})_{\rm SM}$ for various Higgs search channels, $\sigma_{\mu_i}$ is the experimental precision on a particular channel, and the index $i$ runs over all the Higgs search channels in~\autoref{tab:mu_precision}. For the future $\mu_i^{\rm \rm obs}$, we take them to 1. For the future $\delta_{m_h}$, since the present experimental uncertainty, $\sigma_{m_h}^{\rm exp}=0.17 $ GeV \cite{ParticleDataGroup:2020ssz}, we suppose that $\delta_{m_h}$ will be dominated  by the theoretical uncertainty, to be 1 GeV.\footnote{Our study results show a small difference between $\delta_{m_h}$ = 1 GeV and 3 GeV.}~For the $Z$-pole observables, $ X_i=(\Delta S\,, \Delta T\,, \Delta U)_{\rm 2HDM}$ are the 2HDM predicted values, and  $\hat{X}_i=(\Delta S\,, \Delta T\,, \Delta U)$ are the best-fit central values for current measurements (or zero for future measurements)\footnote{Here we only consider effects of $S, T$ and $U$ parameters. There are other paramters discussed such $U, V, W$ \cite{Maksymyk:1993zm} are not included in our study.}.  The $\sigma_{ij}$ are the components of the error matrix, $\sigma_{ij}^2\equiv \sigma_i \rho_{ij} \sigma_j$ where $\sigma_i$ and the correlation matrix components $\rho_{ij} $ are given in~\autoref{tab:STU}.

\begin{table}[tb]
\centering
\resizebox{\textwidth}{!}{
  \begin{tabular}{|l|c|c|c|c|c|c|c|c|c|c|c|c|c|c|c|c|c|c|c|c|}
   \hline
    & \multicolumn{4}{c|}{Current ($1.7 \times 10^{7}\ Z$'s)}& \multicolumn{4}{c|}{CEPC ($10^{10}Z$'s)}& \multicolumn{4}{c|}{FCC-ee ($7\times 10^{11}Z$'s)}&\multicolumn{4}{c|}{ILC ($10^{9}Z$'s)} \\
   \hline
   \multirow{2}{*}{}
   &\multirow{2}{*}{$\sigma$} &\multicolumn{3}{c|}{correlation}
   &{$\sigma$} &\multicolumn{3}{c|}{correlation}
   &{$\sigma$} &\multicolumn{3}{c|}{correlation}
   &{$\sigma$} &\multicolumn{3}{c|}{correlation} \\
   \cline{3-5}\cline{7-9}\cline{11-13}\cline{15-17}
   &&$S$&$T$&$U$&($10^{-2}$)&$S$&$T$&$U$&($10^{-2}$)&$S$&$T$&$U$&($10^{-2}$)&$S$&$T$&$U$\\
   \hline
   $S$& $0.04 \pm 0.11$& 1 & 0.92 & $-0.68$ & $2.46$  & 1     & 0.862       & $-0.373$ &   $0.67$    &  1     &   0.812    &    0.001   &   $3.53$    &   1    &    0.988   & $-0.879$ \\
\hline
   $T$&$0.09\pm 0.14$& $-$ & 1 & $-0.87$ & $2.55$  &  $-$   &  1      &  $-0.735$   &   $0.53$    &   $-$    &    1   &    $-0.097$   &    $4.89$   &   $-$    &   1    &   $-0.909$\\
\hline
   $U$& $-0.02 \pm 0.11$& $-$ & $-$ & 1 &$2.08$  &  $-$   &  $-$     &  1   &   $2.40$    &   $-$    &   $-$    &    1   &  $3.76$     &   $-$    &   $-$    & 1 \\
   \hline
  \end{tabular}
  }
  \caption{Estimated $S$, $T$, and $U$ parameter ranges and correlation matrices $\rho_{ij}$  from $Z$-pole precision measurements, mostly from LEP-I~\cite{ALEPH:2005ab} and future lepton colliders such as CEPC~\cite{CEPC-SPPCStudyGroup:2015csa}, FCC-ee~\cite{Gomez-Ceballos:2013zzn} and ILC ~\cite{Asner:2013psa}. {\textsf{Gfitter}} package~\cite{Baak:2014ora} is used to obtain these constraints.  }
\label{tab:STU}
\end{table}

%%%%%%%%%%%%%%%%%%%%%%%%%%%%%%%%%%%%%%%%%%%%%%%%%%%%%%%%

% ----------------------------------------------
\section{Study results}
\label{sec:results}

\begin{table}[t]
\centering
\begin{tabular}{ c|c|c|c|c|c } 
 \hline
 &$\lambda_1$ & $\lambda_2$ & $\lambda_3$ &$\lambda_4$ &$\lambda_5$ \\ 
 \hline
 Region & $(0,\,5.5)$ & $(0,\,1)$ & $(-1.8,\,1.5)$ & $(-2,\,2)$ & $(-1.4,\,1.2)$ \\
 \hline
\end{tabular}
\caption{Allowed region in the generic basis $\lambda_{1-5}$ at the 95\% confidence level.}
\label{tab:lambda1-5}
\end{table}

%%% combined couplings
\begin{table*}[t]
    \centering
    \newcolumntype{A}{ >{\centering\arraybackslash} m{.005\linewidth}}
    \newcolumntype{B}{ >{\centering\arraybackslash} m{.175\linewidth} }
    \newcolumntype{C}{ >{\centering\arraybackslash} m{.175\linewidth} }
    \newcolumntype{D}{ >{\centering\arraybackslash} m{.175\linewidth} }
    \newcolumntype{E}{ >{\centering\arraybackslash} m{.175\linewidth} }
    \newcolumntype{F}{ >{\centering\arraybackslash} m{.175\linewidth} }
    \begin{tabular}{ A B C D E F }
        \rotatebox{90}{$\lambda_1$}
        & \hspace{0.04\linewidth}\includegraphics[width=0.92\linewidth]{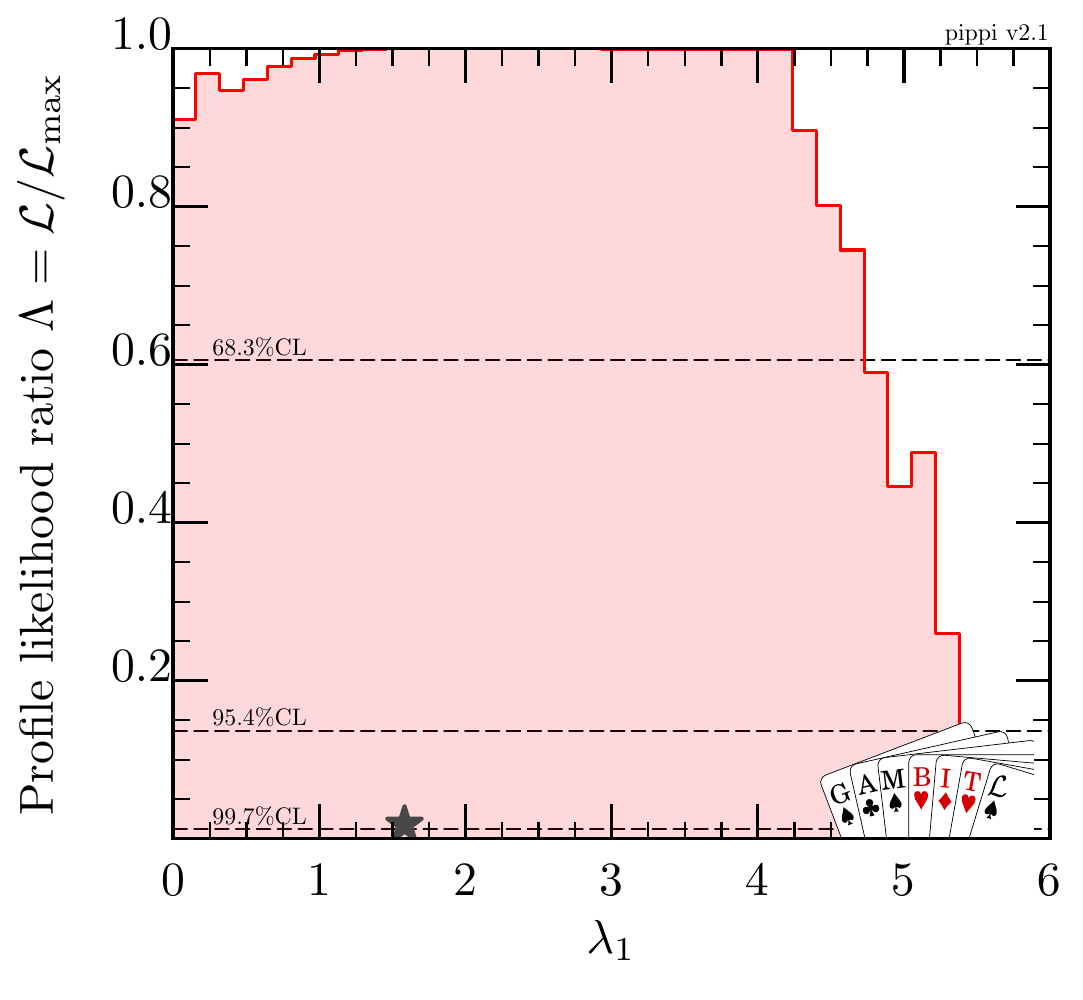} \hspace{0.04\linewidth}
        & & & & \\
        \rotatebox{90}{$\lambda_2$}
        & \includegraphics[width=\linewidth]{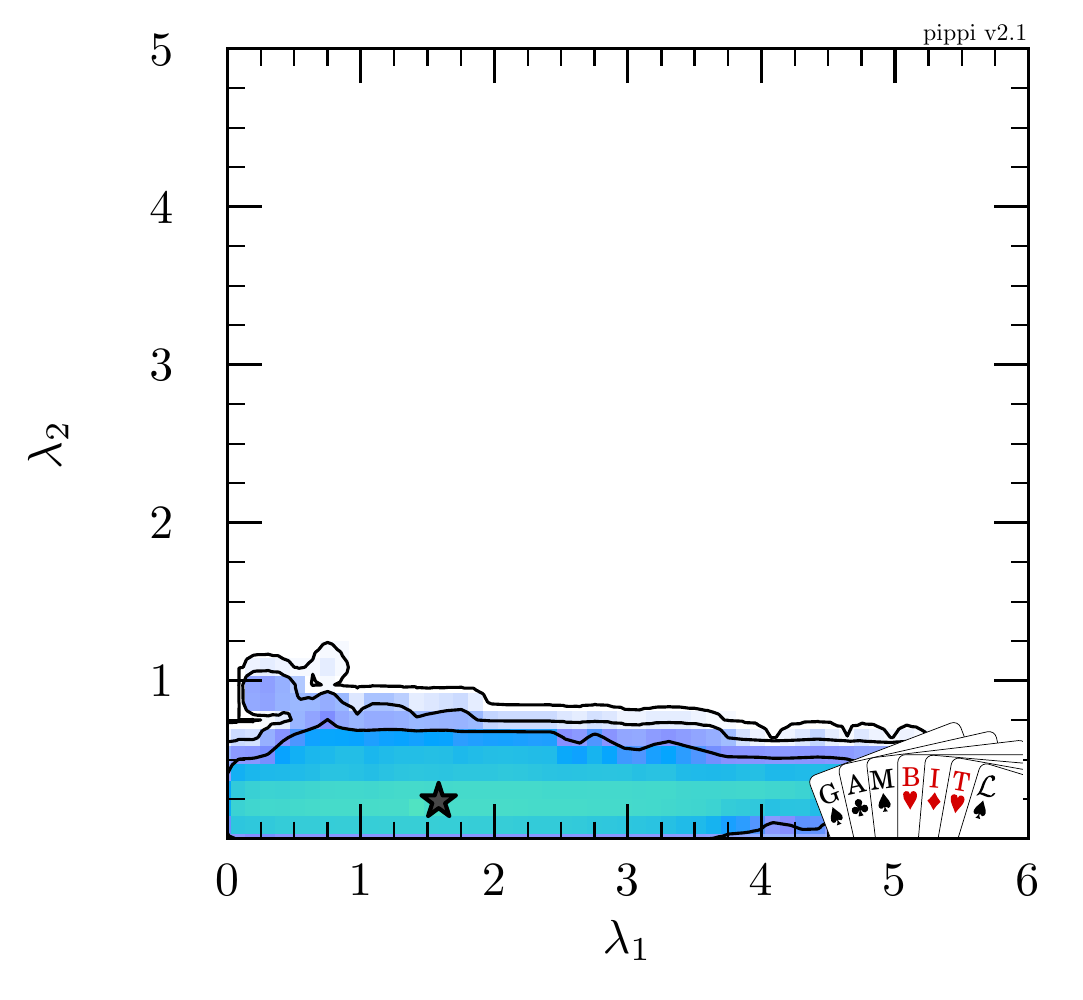}
        & \hspace{0.04\linewidth}\includegraphics[width=0.92\linewidth]{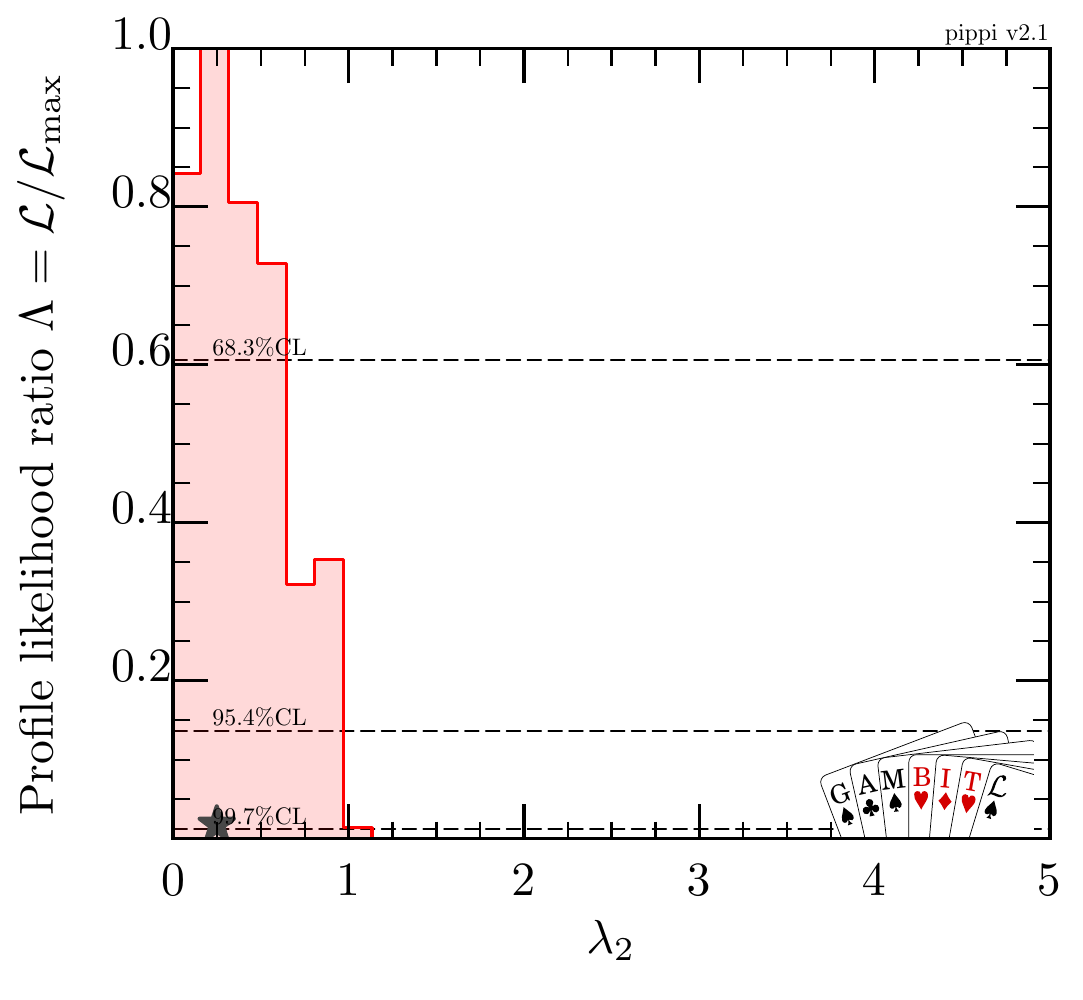} \hspace{0.04\linewidth}
        & & & \\
        \rotatebox{90}{$\lambda_3$}
        & \includegraphics[width=\linewidth]{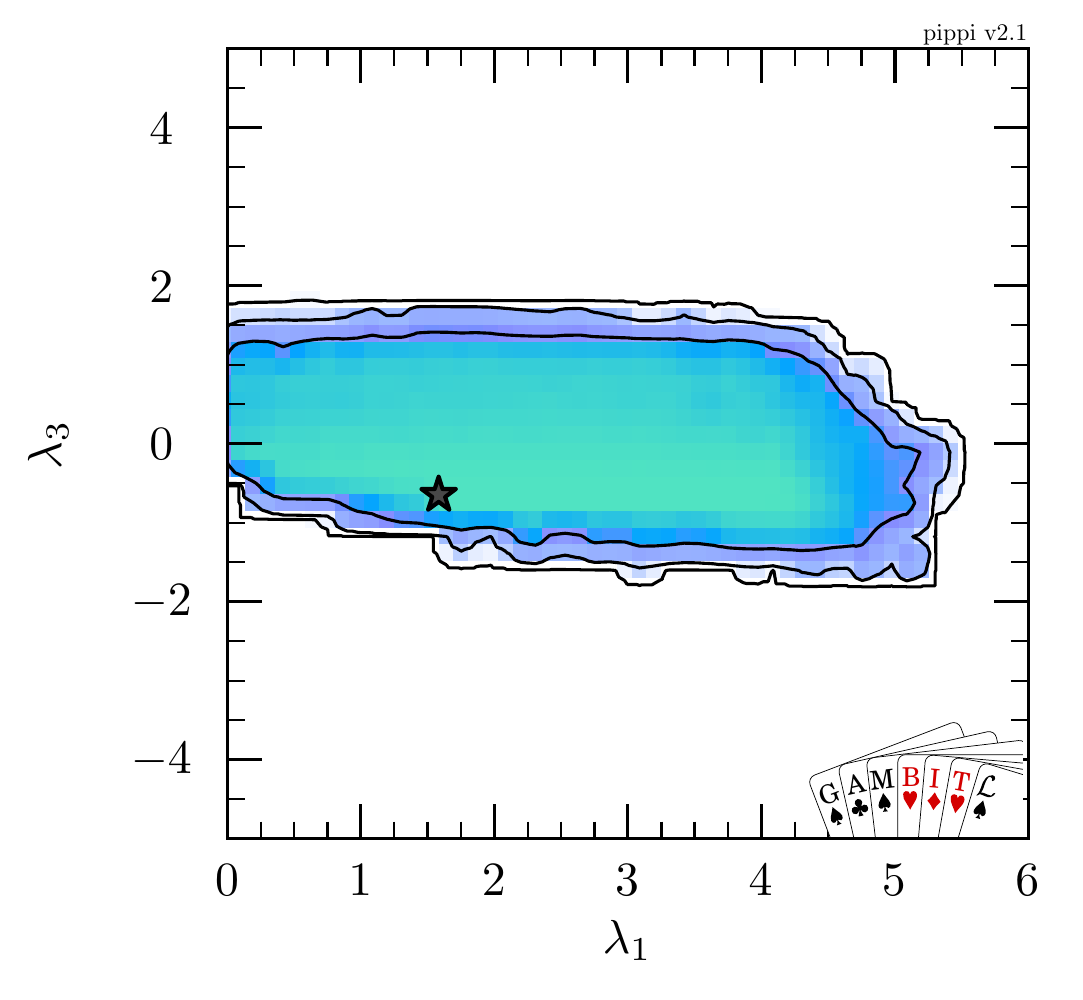}
        & \includegraphics[width=\linewidth]{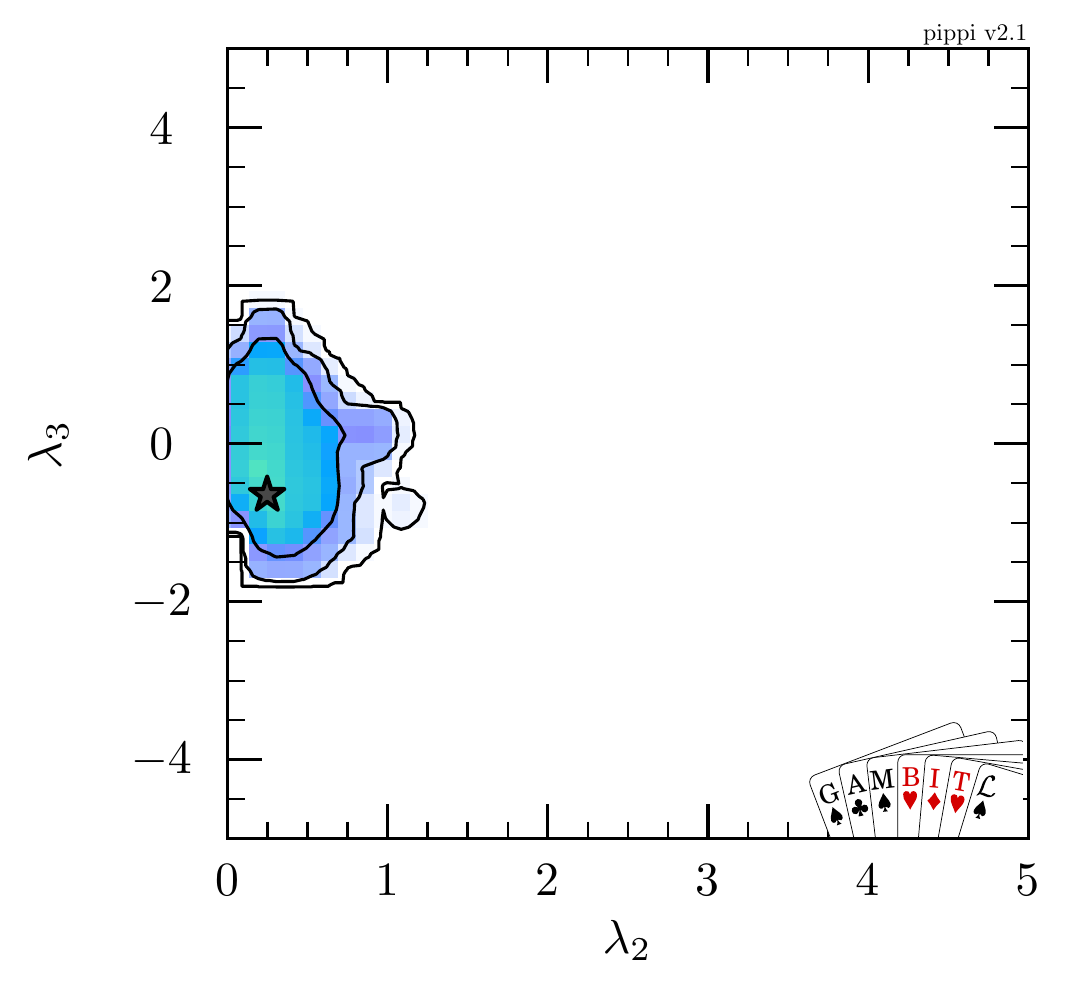}
        & \hspace{0.04\linewidth}\includegraphics[width=0.92\linewidth]{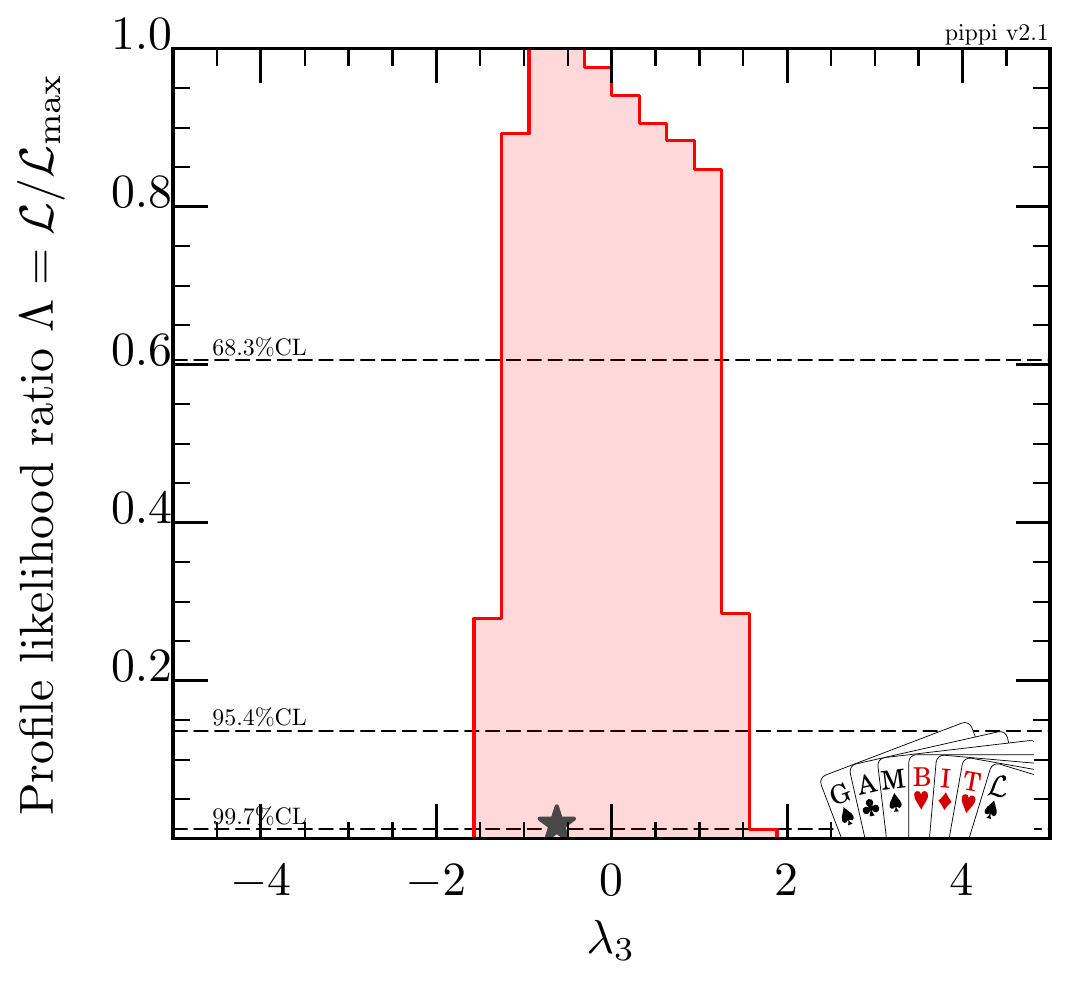} \hspace{0.04\linewidth}
        & & \\
        \rotatebox{90}{$\lambda_4$}
        & \includegraphics[width=\linewidth]{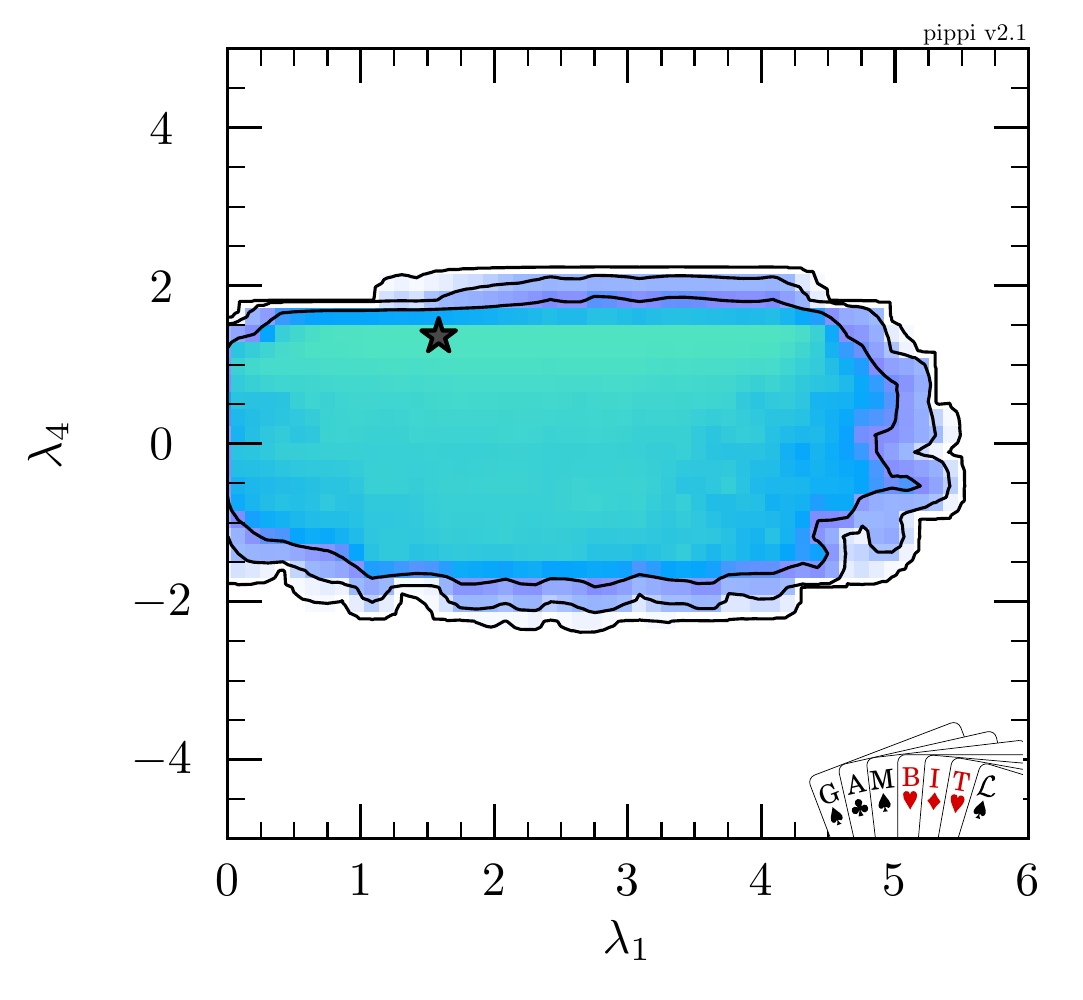}
        & \includegraphics[width=\linewidth]{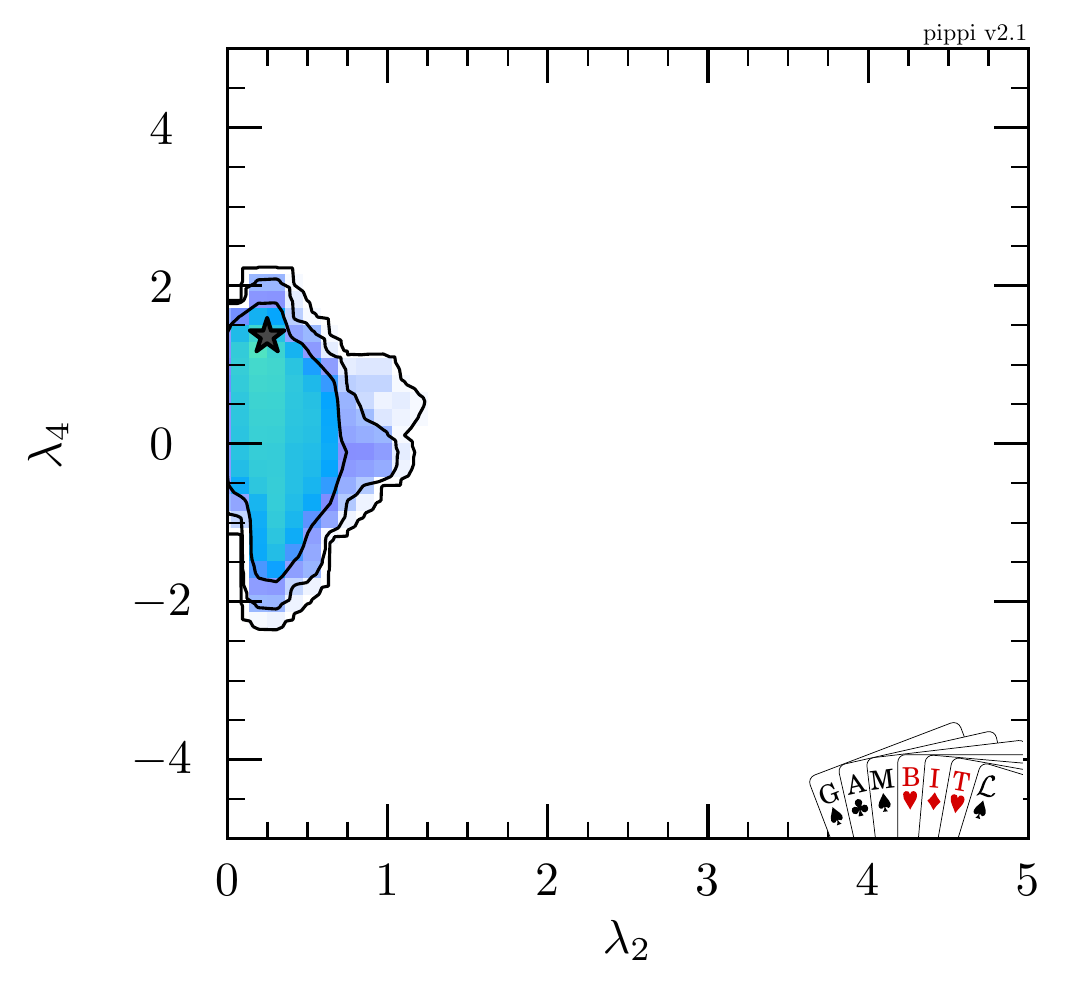}
        & \includegraphics[width=\linewidth]{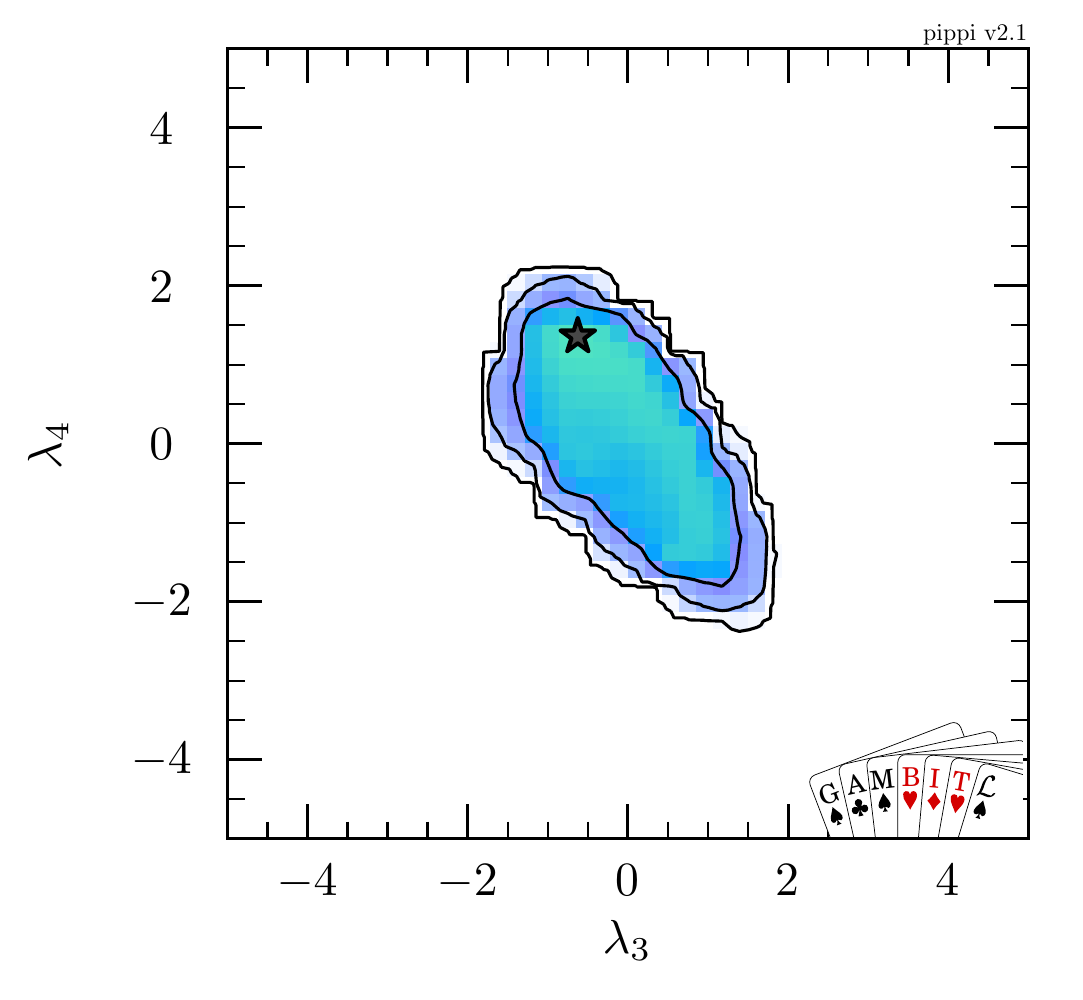}
        & \hspace{0.04\linewidth}\includegraphics[width=0.92\linewidth]{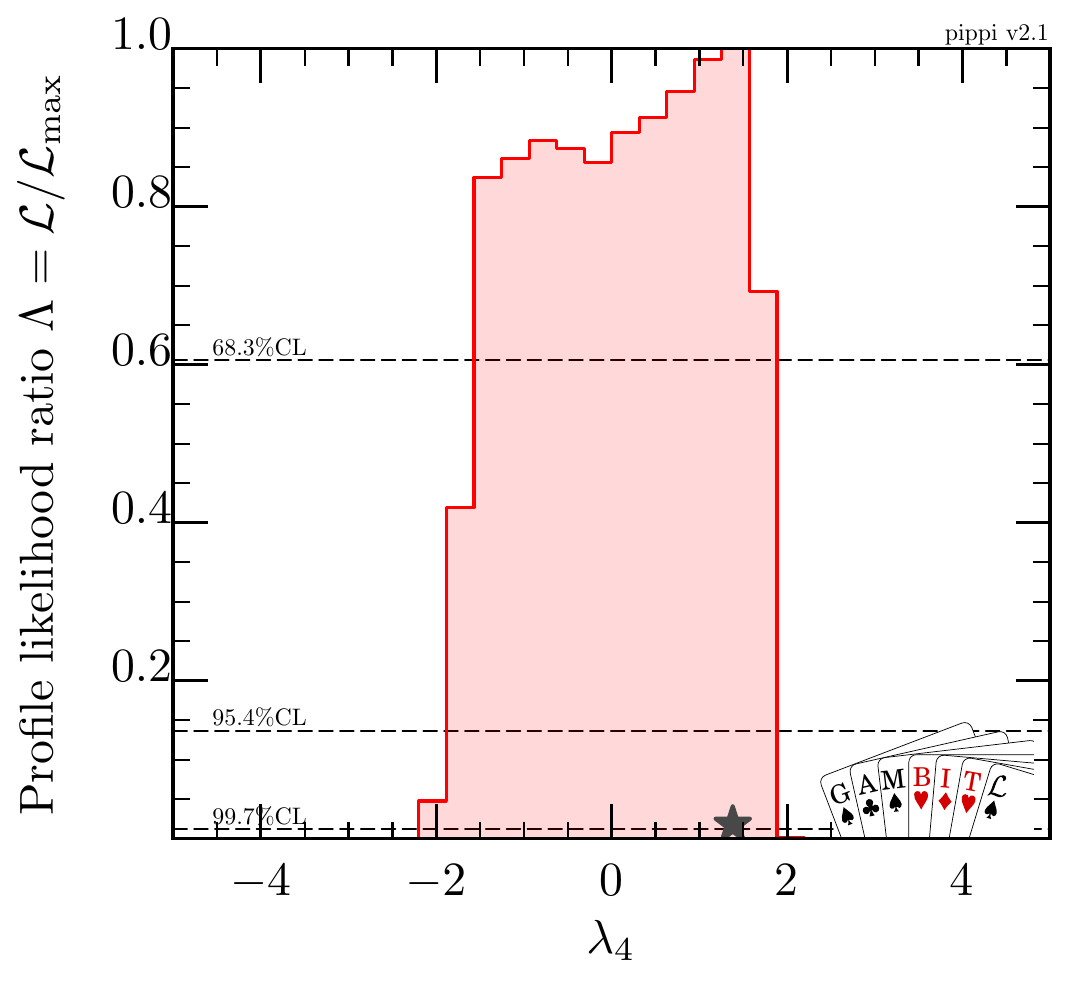} \hspace{0.04\linewidth}
        & \\
        \rotatebox{90}{$\lambda_5$}
        & \includegraphics[width=\linewidth]{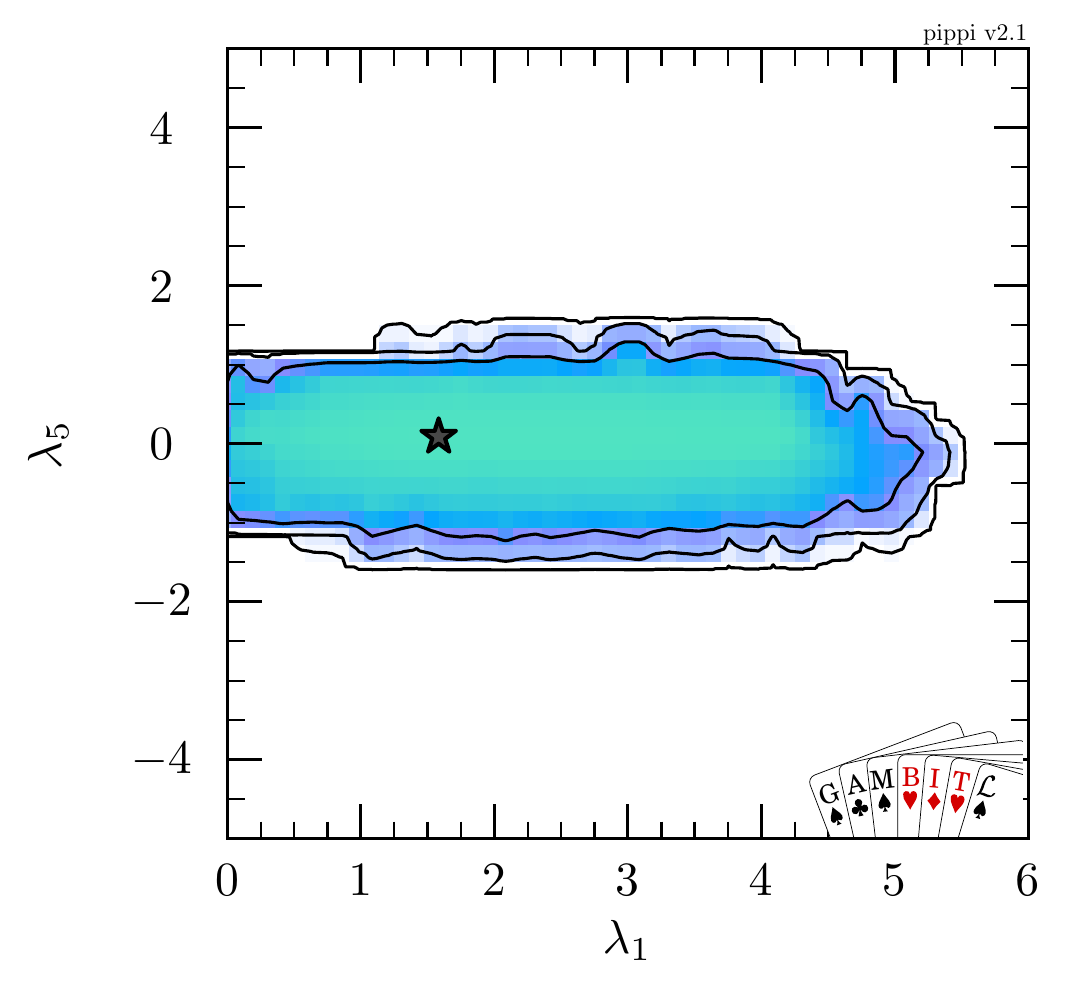}
        & \includegraphics[width=\linewidth]{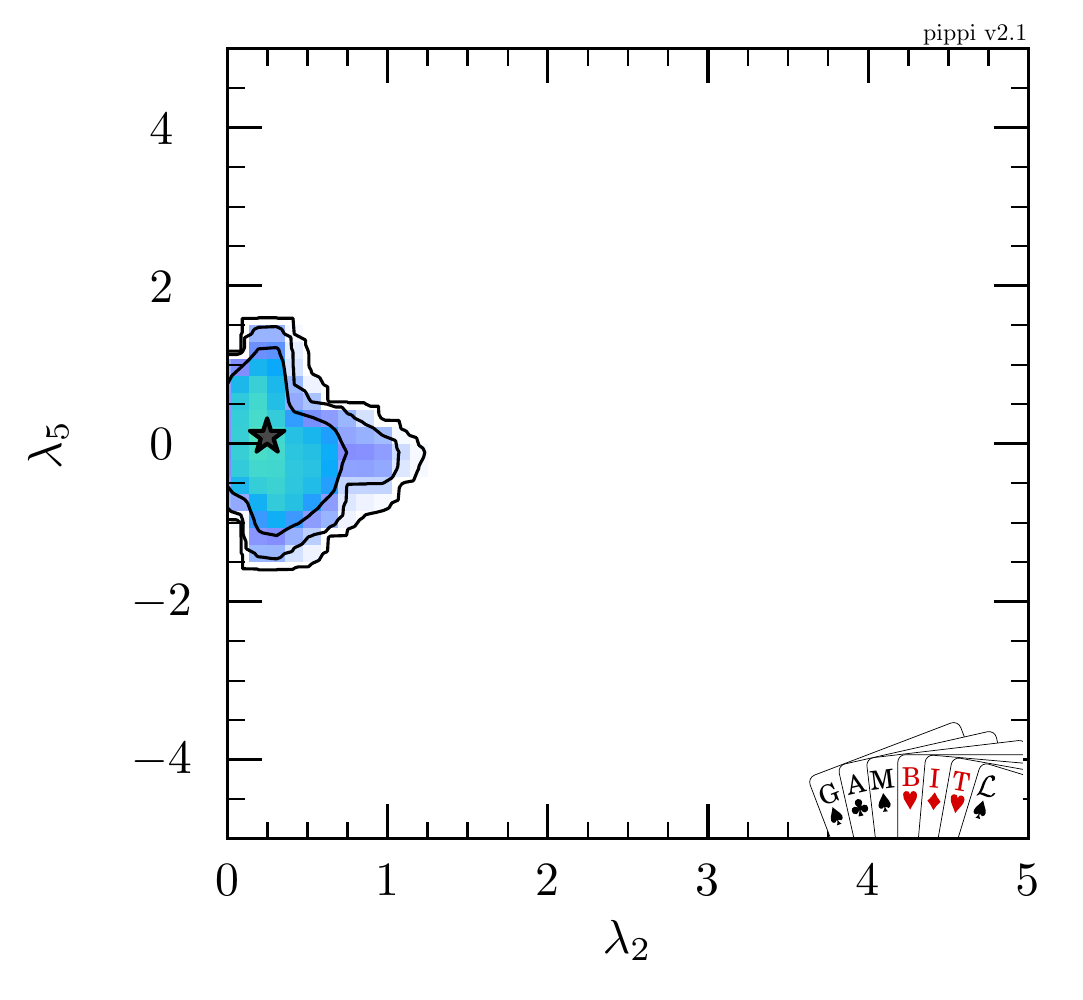}
        & \includegraphics[width=\linewidth]{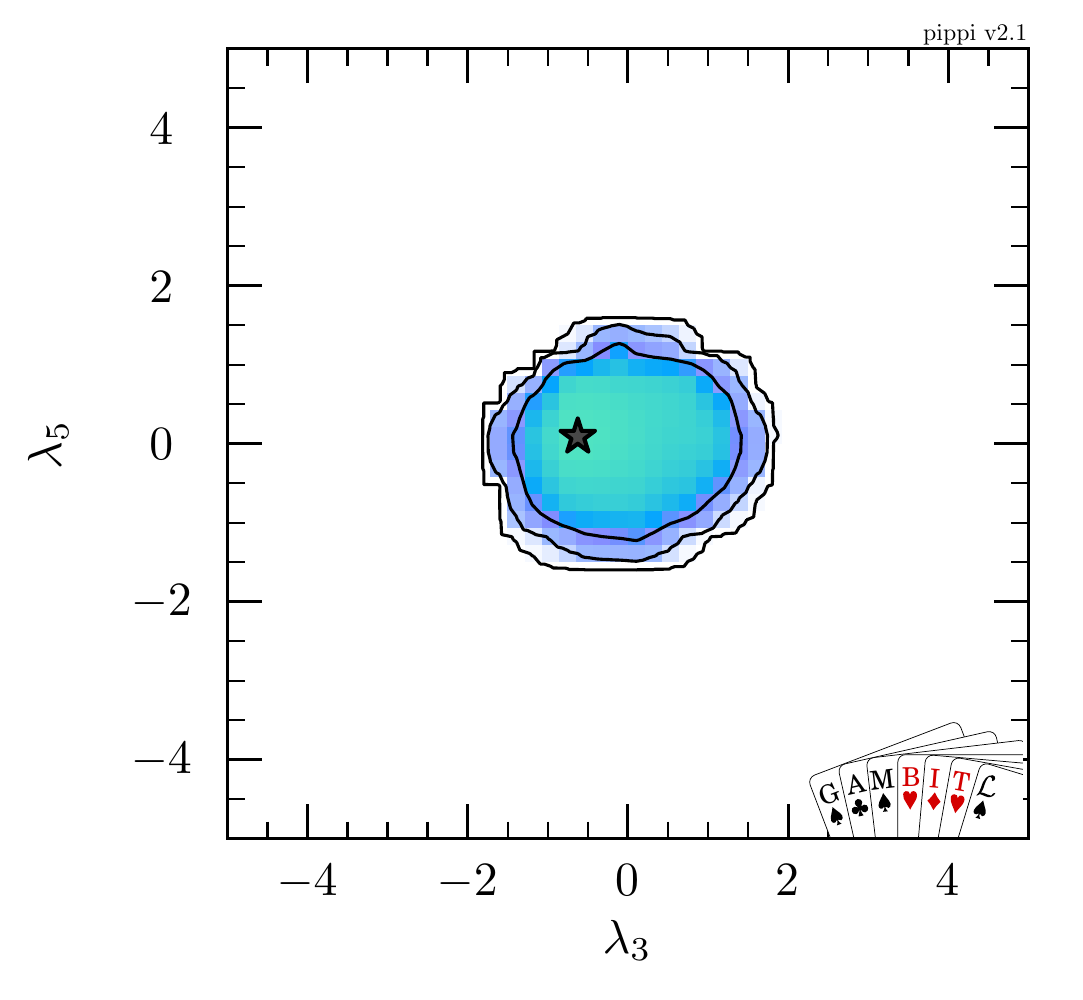}
        & \includegraphics[width=\linewidth]{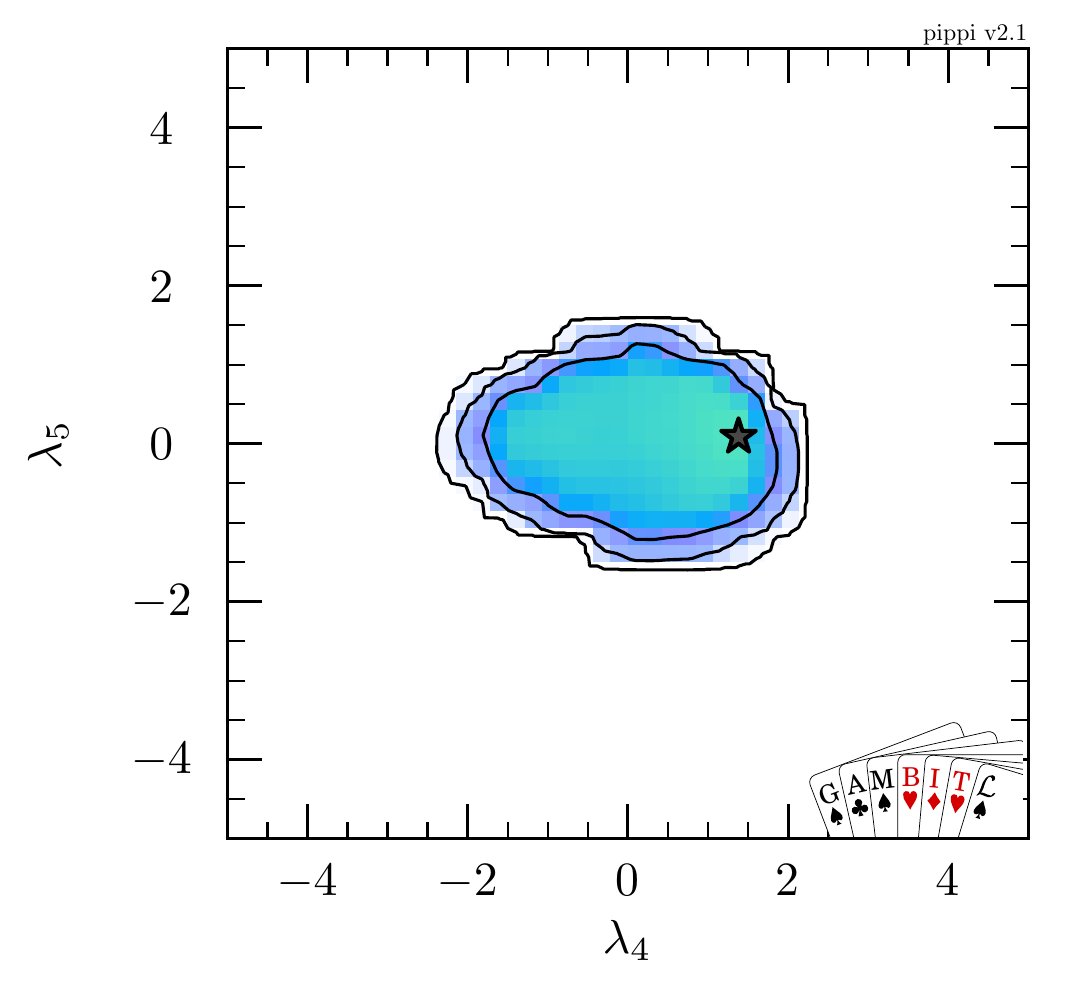}
        & \hspace{0.04\linewidth}\includegraphics[width=0.92\linewidth]{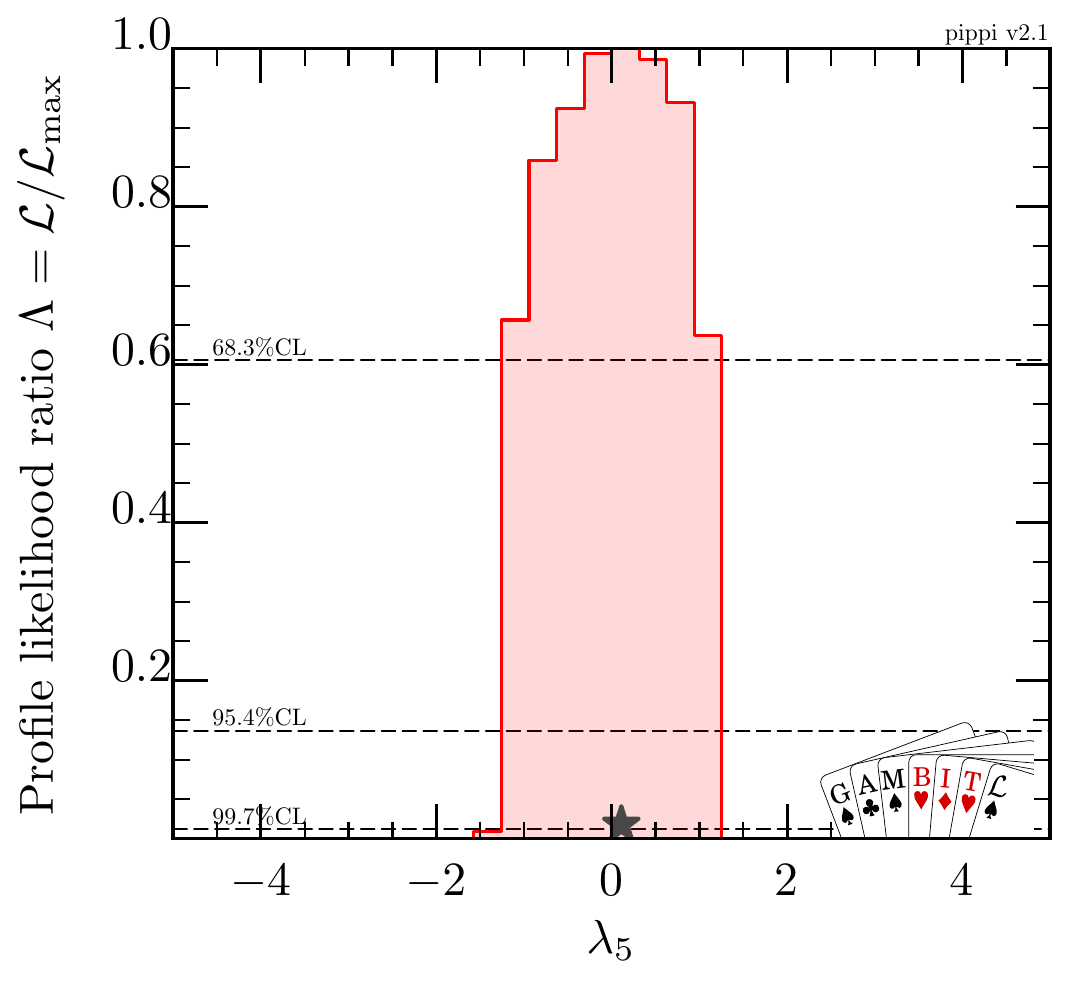} \hspace{0.04\linewidth}
        \\
        & $\lambda_1$ & $\lambda_2$ & $\lambda_3$ & $\lambda_4$ & $\lambda_5$ \\
    \end{tabular}
    \captionof{figure}{Combined 1D and 2D profile likelihood distributions for the couplings in the generic basis, with marked boundaries for the 1$\sigma$, 2$\sigma$ and 3$\sigma$ confidence regions. The global fit includes theoretical constraints and current collider Higgs, electroweak precision and flavour constraints.} % \usepackage{caption}
    \label{fig:combined_couplings}
\end{table*}

In Fig.~\ref{fig:combined_couplings}, we show the 1D and 2D profile likelihood distributions for the couplings in the generic basis, with marked boundaries for the 1$\sigma$, 2$\sigma$ and 3$\sigma$ confidence regions. These results include all of the latest relevant constraints, including theoretical constraints (unitarity, perturbativity and vacuum stability), Higgs searches at colliders based on latest version of  \textsf{HiggsBounds-5.3.2} \cite{Bechtle:2008jh,Bechtle:2015pma} and \textsf{HiggsSignals-2.2.3}~\cite{Bechtle:2013xfa}, electroweak physics and flavour constraints. A more detailed description will be provided in the forthcoming \textsf{GAMBIT} paper~\cite{gambit_Type-II_2HDM}. Generally speaking, there are still large allowed regions in the general basis as summarized in~\autoref{tab:lambda1-5}, and the regions change little after including hypothetical future results.

\begin{figure*}[ht]
  \centering
 \includegraphics[width=0.45\linewidth]{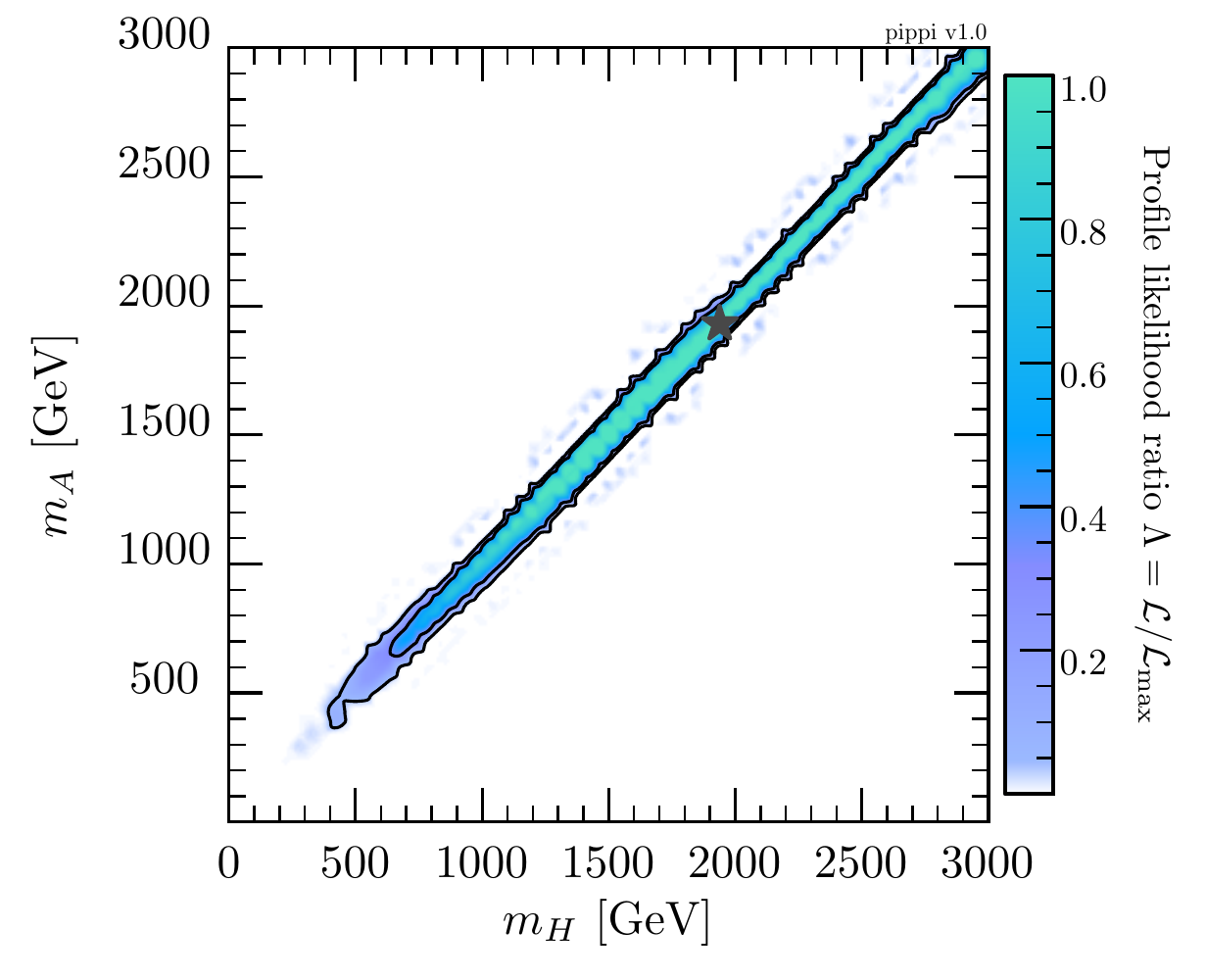}
 \includegraphics[width=0.52\linewidth]{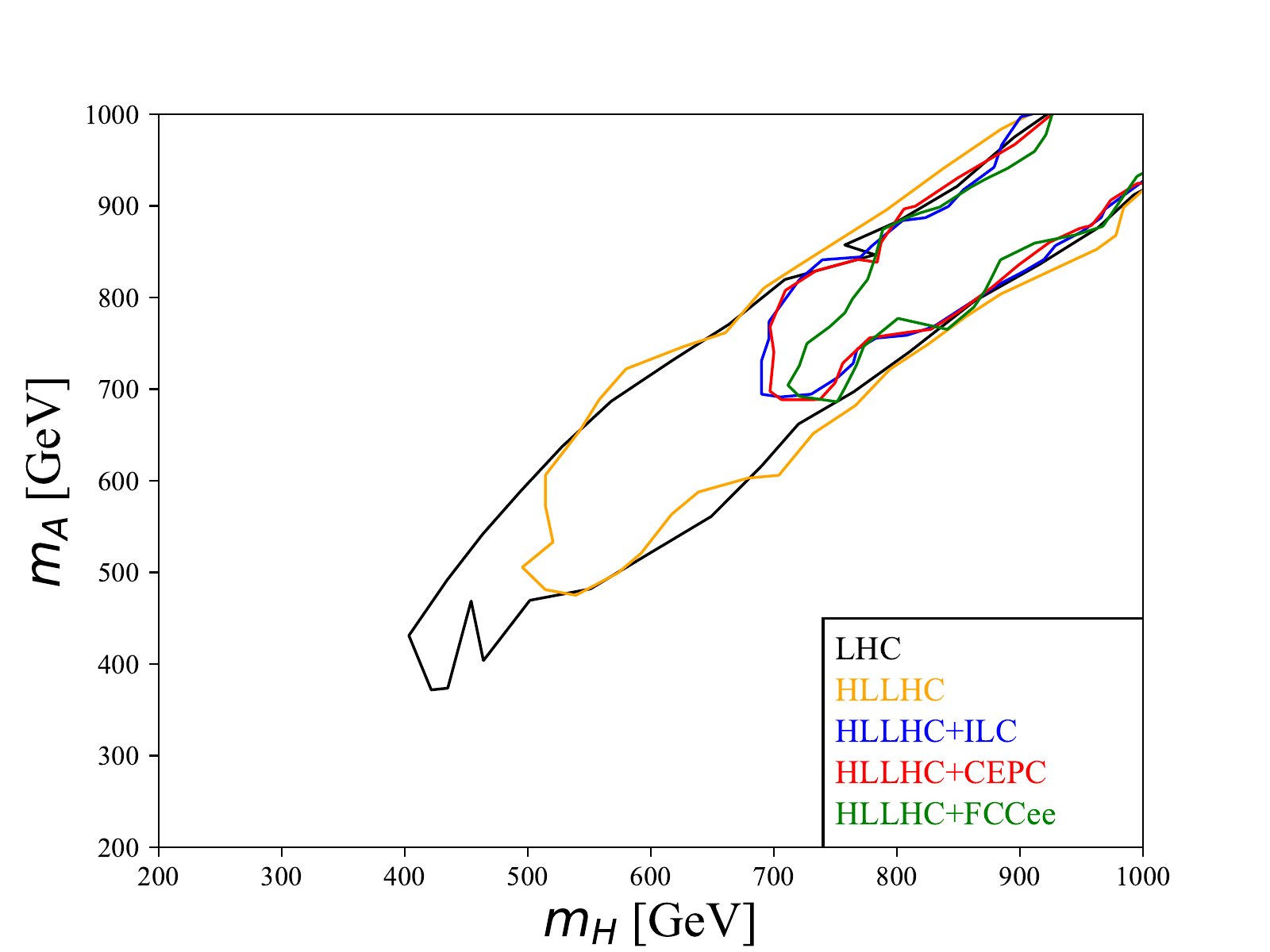}
\includegraphics[width=0.45\linewidth]{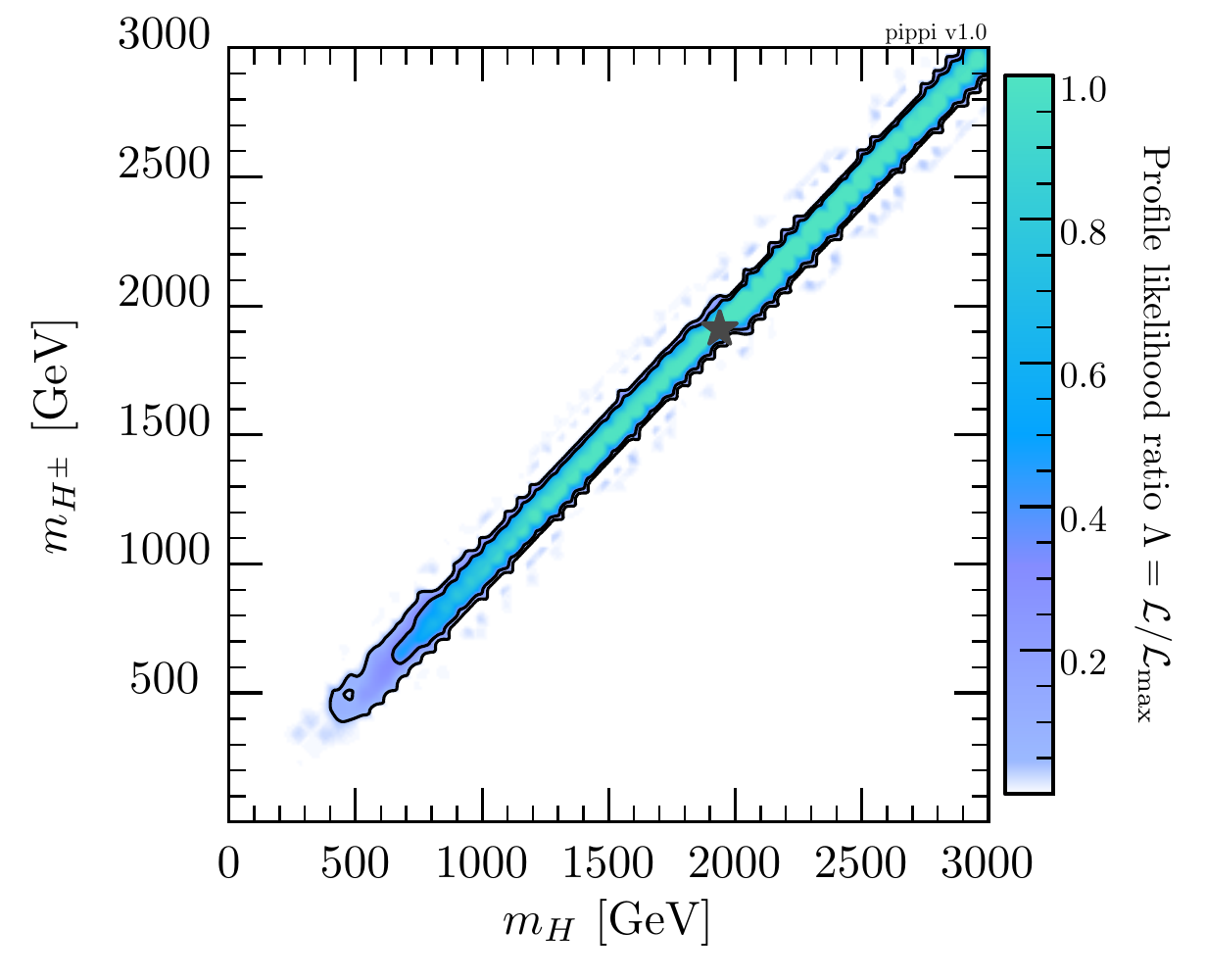}
\includegraphics[width=0.52\linewidth]{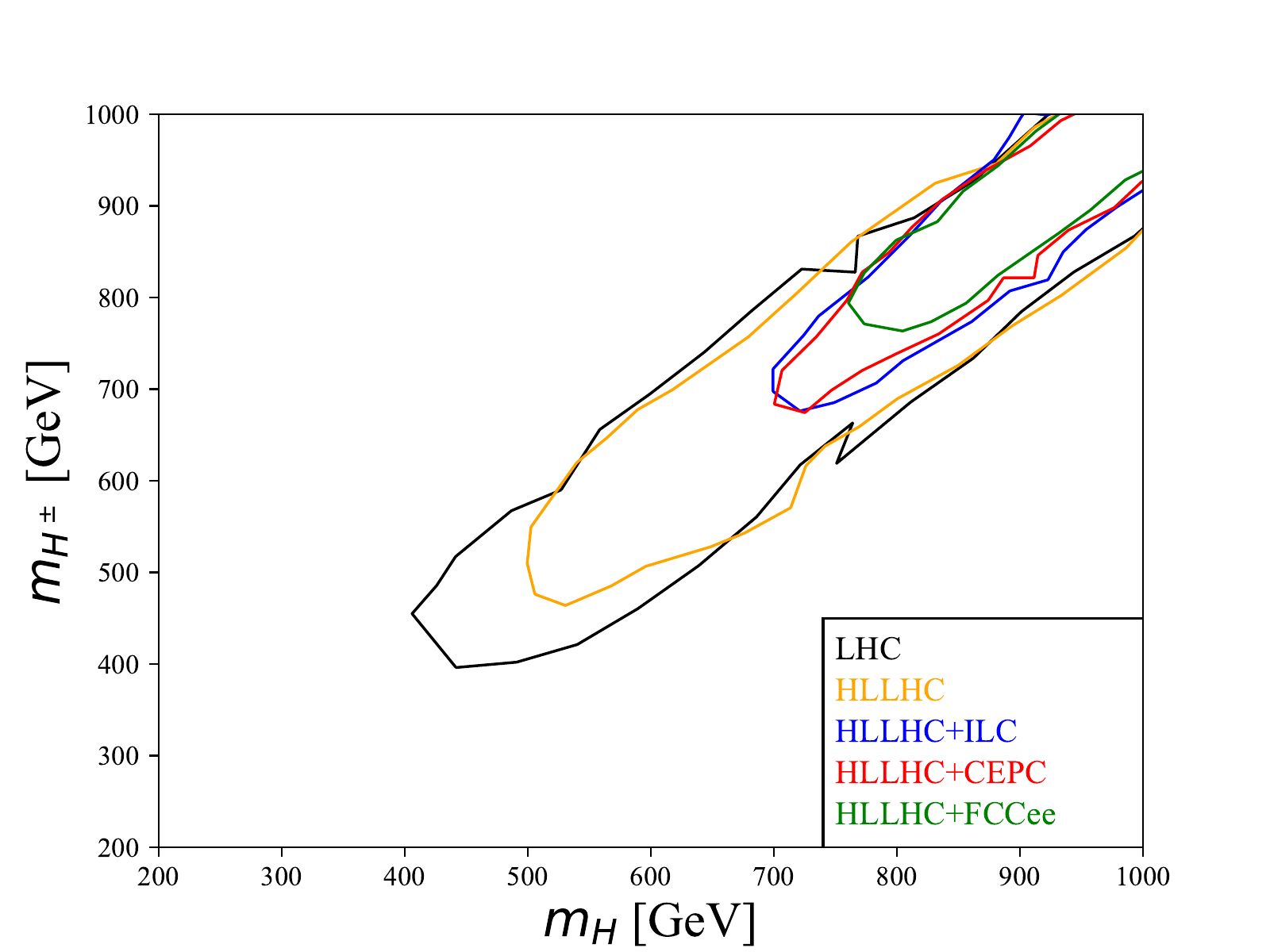}
  \includegraphics[width=0.45\linewidth]{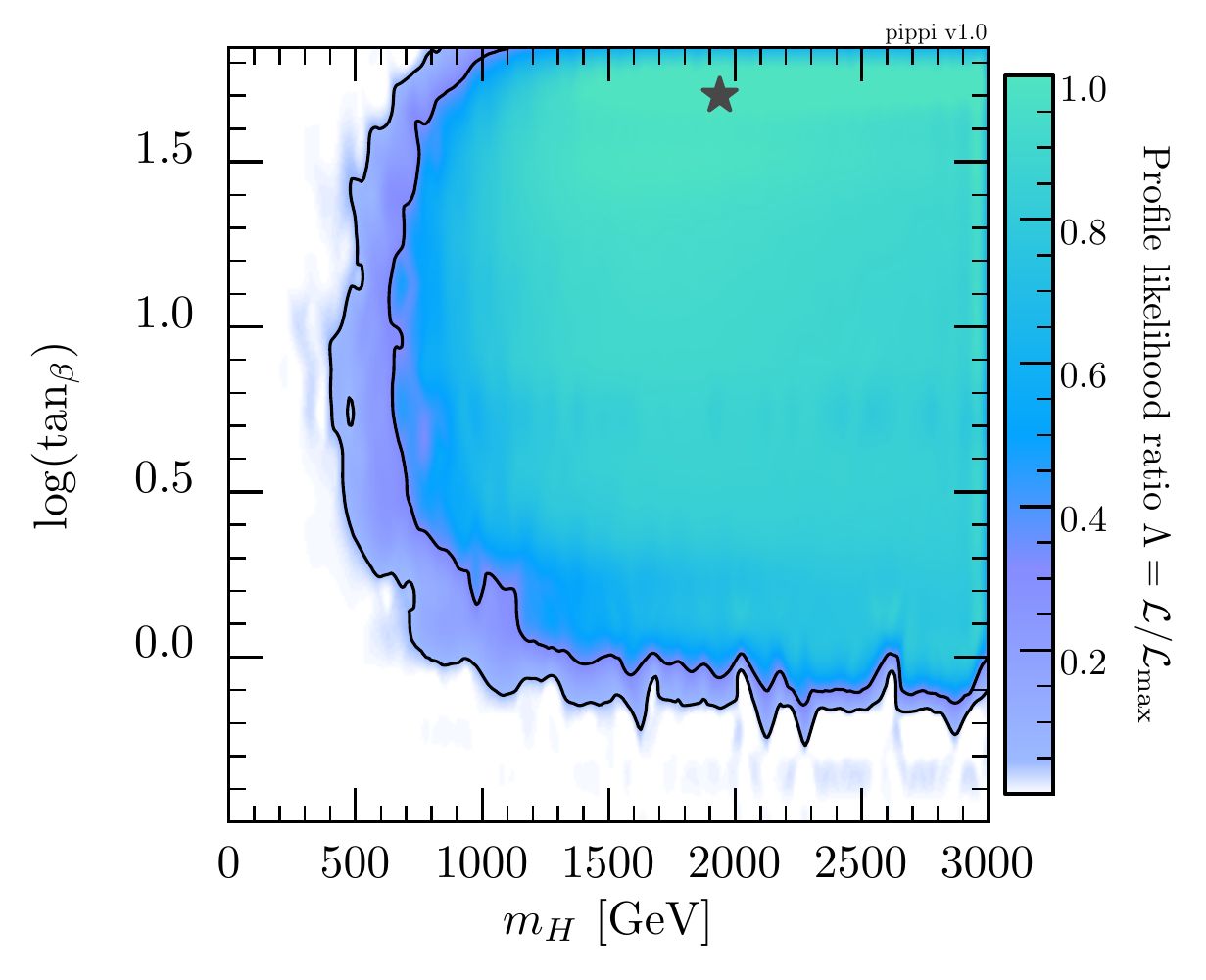}
\includegraphics[width=0.52\linewidth]{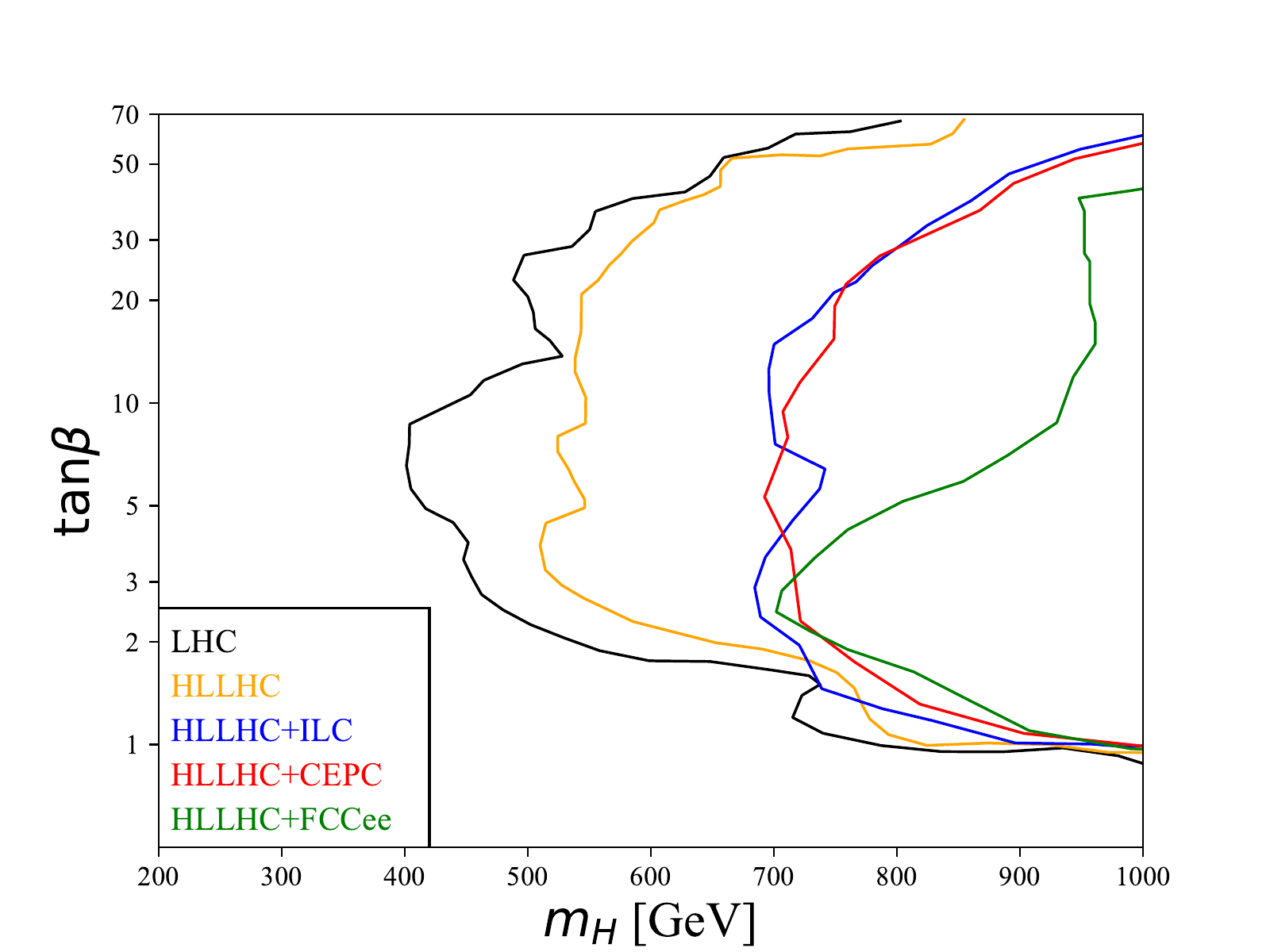}
  \caption{Left panel: Global fit results showing the 1$\sigma$ and 2$\sigma$ regions in the $m_{H}$-$m_A$ plane (top), $m_{H}$-$m_{H^\pm}$ plane (middle) and $m_H$-$\tan\beta$ plane (bottom) based on current measurements and theoretical constraints. Right panel: Comparison of current $2\sigma$ constraints (black) with those arising from the inclusion of future precision measurements, including those from the HL-LHC (orange), HL-LHC + CEPC (red), HL-LHC + ILC (blue), and HL-LHC + FCC-ee (green). 
  }
  \label{fig:mH_mC}
\end{figure*}

In Fig.~\ref{fig:mH_mC}, we show the global fit results in the $m_{H}$-$m_A$ plane (top), $m_{H}$-$m_{H^\pm}$ plane (middle) and $m_H$-$\tan\beta$ plane (bottom). The left panels show the global fit results with current data and theoretical constraints. Generally speaking, there are lower limits on the heavy scalar masses of approximately 400 GeV, which mainly comes from the $Z$-pole precision measurements and flavour physics. There is also a lower limit on $\tan\beta$ that arises mainly from flavor physics. 
The right panels compare the current 2$\sigma$ confidence regions (black) with those arising from the inclusion of future precision measurements, including those from the HL-LHC (orange), HL-LHC + CEPC (red), HL-LHC + ILC (blue), and HL-LHC + FCC-ee (green).~The inclusion of HL-LHC precision measurements pushes the lower limit on the scalar mass up to 500 GeV, and these rise further up to 700 GeV after including constraints from future lepton colliders.

\begin{figure*}[t]
  \centering
 \includegraphics[width=0.45\linewidth]{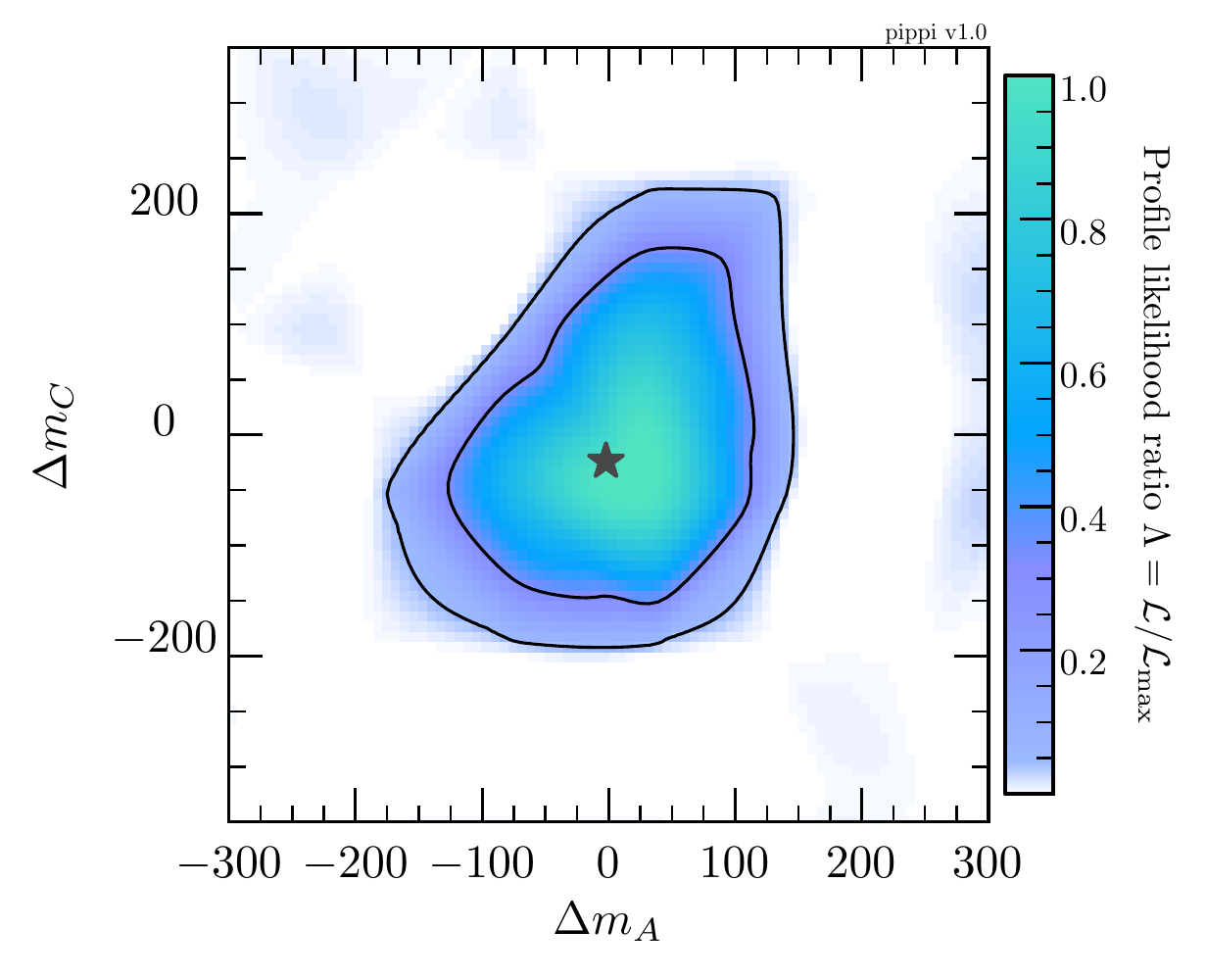}
 \includegraphics[width=0.52\linewidth]{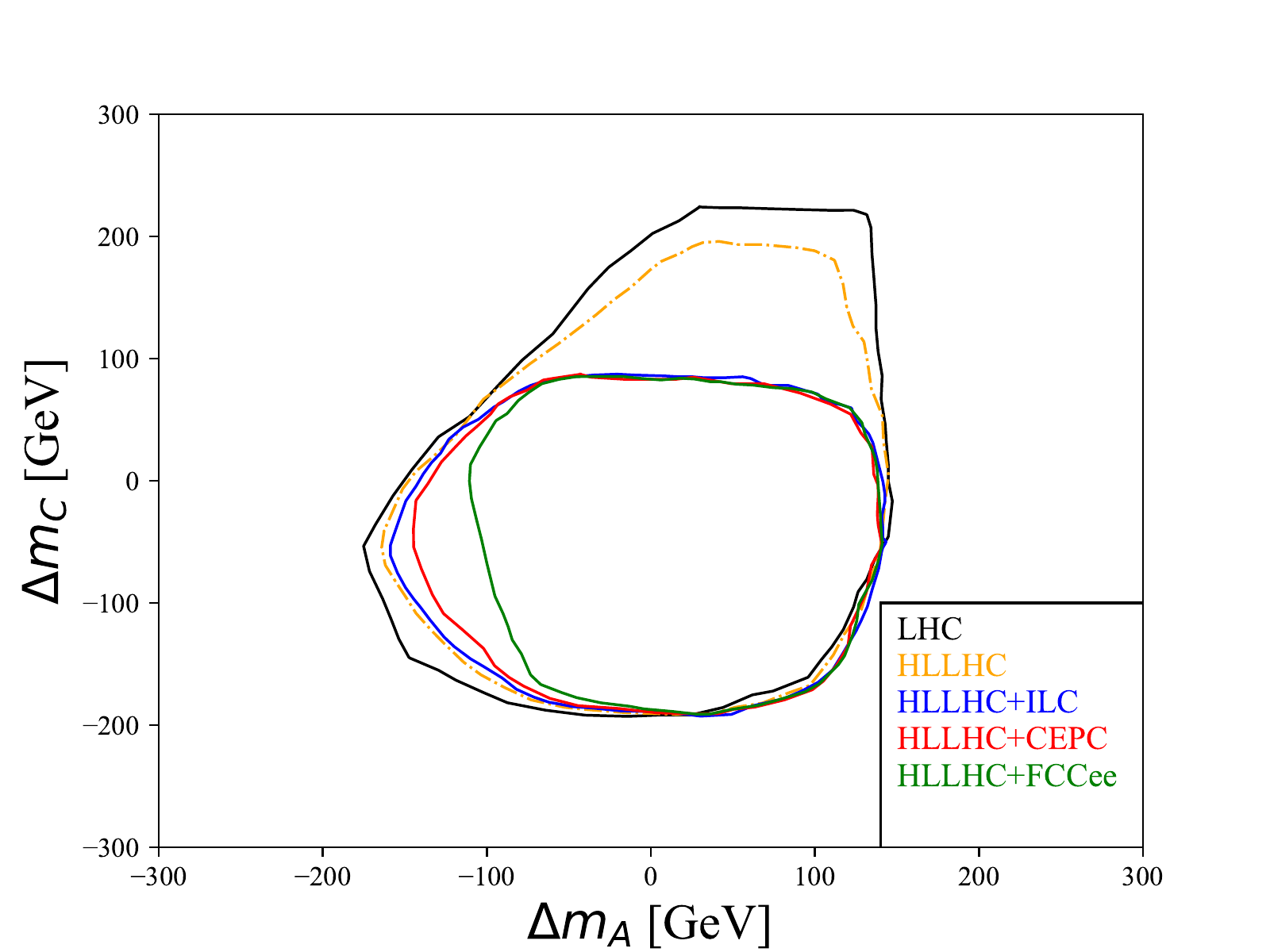}
  \caption{Left panel: The 1$\sigma$ and 2$\sigma$ allowed regions in $\Delta m_{A}- \Delta m_C$ plane, based on current measurements and theoretical constraints. Right panel: Comparison  of the $2\sigma$ region allowed by current constraints (black) with those arising from the inclusion of future precision measurements, including those from the HL-LHC (orange), HL-LHC+CEPC (red), HL-LHC + ILC(blue), and HL-LHC + FCC-ee (green). 
  }
  \label{fig:dmadmc}
\end{figure*}

Finally in Fig.~\ref{fig:dmadmc}, we show the 1$\sigma$ and 2$\sigma$ regions allowed by current measurements and theoretical constraints in the plane of $\Delta m_A = m_A-m_H$ vs $\Delta m_C = m_{H^\pm}-m_H$ (left panel). The right panel compares the 2$\sigma$ region (black) with those arising from the inclusion of future precision measurements including those from the HL-LHC (orange), HL-LHC + CEPC (red), HL-LHC + ILC (blue), and HL-LHC + FCC-ee (green). At present, $\Delta m_A$ and $\Delta m_C$ are limited to the range $(-200,150)$\,GeV and $(-200,250)$\,GeV respectively, both of which will be reduced to $(-200,100)$\,GeV in the future. 

We also notice that the expected limit from the FCC-ee proposal is a little stronger than those arising from the ILC and CEPC proposals. This is mainly because the currently proposed FCC-ee will have a larger luminosity around 250 GeV, which is about 2 times that of CEPC. At the same time, it will also produce more $Z$ bosons than the other two lepton colliders. 

To summarise, after comparing global fit results with current and hypothetical future data, we find that:
\begin{enumerate}
    \item In the generic coupling basis, there is not much ability to constrain $\lambda_{1-5}$ from future Higgs mass, coupling, and $Z$-pole precision measurements.
    \item The same is not true when considering the physical masses of the heavy scalars. Currently, the masses $m_{H,A,H^\pm}$ have a lower limit of approximately 400 GeV, which is increased to 500 and 700\,GeV after including hypothetical results from the HL-LHC and HL-LHC + lepton colliders. 
    \item There are strong constraints on the mass splittings $\Delta m_A = m_A-m_H$ and $\Delta m_C = m_{H^\pm}-m_H$ from Higgs and $Z$-pole precision measurements, which are currently limited to the ranges $(-200,150)$\,GeV and $(-200,250)$\,GeV. Future precision measurements will shrink the allowed range on $\Delta m_C$ to $(-200,100)$\,GeV.
\end{enumerate}

%
% \clearpage
\section{Acknowledgement}
We acknowledge helpful discussions with the GAMBIT community.
\bibliographystyle{JHEP}
\bibliography{references}

\end{document}